\documentclass[aps,prl,twocolumn,superscriptaddress,showpacs]{revtex4-1}
\usepackage{amsmath,amssymb,graphics,epsfig,epstopdf,color,verbatim,ulem,braket,tabularx}
\usepackage[colorlinks,linkcolor=blue,citecolor=blue,urlcolor=blue]{hyperref}

\begin{document}

\title{Dynamical Generation of Topological Masses in Dirac Fermions}

\author{Yuan-Yao He}
\address{Department of Physics, Renmin University of China, Beijing 100872, China}
\author{Xiao Yan Xu}
\address{Beijing National Laboratory for Condensed Matter Physics, and Institute of Physics, Chinese Academy of Sciences, Beijing 100190, China}
\address{School of Physical Sciences, University of Chinese Academy of Sciences, Beijing, 100190, China}
\author{Kai Sun}
\address{Physics Department, University of Michigan, Ann Arbor, MI 48109, USA}
\author{Fakher F. Assaad}
\address{Institut f\"ur Theoretische Physik und Astrophysik, Universit\"at W\"urzburg, 97074 W\"urzburg, Germany}
\author{Zi Yang Meng}
\address{Beijing National Laboratory for Condensed Matter Physics, and Institute of Physics, Chinese Academy of Sciences, Beijing 100190, China}
\address{School of Physical Sciences, University of Chinese Academy of Sciences, Beijing, 100190, China}
\author{Zhong-Yi Lu}
\address{Department of Physics, Renmin University of China, Beijing 100872, China}

\begin{abstract}
We report discovery of a topological Mott insulator in strongly-correlated Dirac semimetals. Such  an interaction-driven topological state  has been theoretically proposed but not yet observed with  unbiased  large scale  numerical simulations.   In our model,  interactions between electrons are  mediated by Ising spins in a transverse field.    The results indicate  that the topological mass term  is dynamically generated  and the resulting quantum phase transition  belongs to the (2+1)D $N=8$ chiral Ising universality class.   These conclusions stem  from large scale sign free quantum Monte Carlo simulations.
\end{abstract}


\date{\today}
\maketitle

\textit{Introduction}. Combination of the richness of quantum many-body effects and the elegance of topological physics~\cite{Klitzing1980,Thouless1982,Haldane1988,Hasan2010,XiaoLiang2011}  has revealed remarkable phenomena and new principles of physics, such as the fractional quantum Hall effect~\cite{Tsui1982,Laughlin1983} and topological order~\cite{XiaoGang1990}. Among these discoveries, one intriguing example is interaction-driven topological states, where strong correlations among particles convert a conventional state of matter into a topological one. One pathway towards such states is to utilize the phenomenon of spontaneous symmetry breaking~\cite{Raghu2008,Sun2009,Herbut2014,Zhang2009,Yu61}, i.e. in a system where nontrivial topological structures are prohibited by symmetry, strong interactions can spontaneously  break symmetry and thus stabilize a topologically nontrivial ground state. As proposed in Ref.~\cite{Raghu2008}, such a phenomenon can arise in a 2D Dirac semimetal (DSM) through a quantum phase transition that breaks spontaneously the time-reversal or the spin rotational symmetry, resulting
in an interaction-driven, quantum-Hall or quantum-spin-Hall (QSH), topological insulator, dubbed topological Mott insulators (TMI).

Although the general principle about TMI has been well understood, finding such a state via unbiased theoretical/numerical methods turns out to be challenging due to the strong coupling nature of the problem and the presence of competing orders. Extensive numerical efforts on interacting DSMs~\cite{Valenzuela2013,Daghofer2014,Huaiming2014,Motruk2015,Capponi2015,Scherer2015} report negative results, suggesting that in all explored parameter regimes, topologically-trivial competing states always have lower energy and thus the proposed TMI states cannot be stabilized. The successful alternative came lately, by substituting the DSM by a semimetal with a quadratic band crossing~\cite{Sun2009,Herbut2014,Sun2012}, an interaction-driven quantum Hall state  is observed numerically~\cite{HanQing2016B}. Furthermore, experimental realization of such scenario has very recently been proposed in functionalized $\alpha$-Fe$_{2}$O$_{3}$ nanosheet~\cite{Liang2017}. However, whether a TMI can emerge from a DSM without the assistance of a quadratic band crossing point, as in the original proposal~\cite{Raghu2008}, still remains an open question. It is also worthwhile to highlight that between the two possible types of TMIs, quantum Hall and quantum spin Hall~\cite{Raghu2008}, only the former has been observed in numerical studies~\cite{HanQing2016B}. Hence, to find a time-reversal invariant TMI is one key objective of this study.

On the other hand, in a seemingly unrelated research area,  recent developments in  sign-problem-free quantum Monte Carlo (QMC) approaches for itinerant fermions coupled to fluctuating bosonic fields open the door to  investigate many intriguing strongly-correlated systems, such as antiferromagnetic fluctuations mediated superconductivity in metals~\cite{Berg1606,Schattner2016B}, nematic quantum critical points in itinerant systems~\cite{Lederer2015,Schattner2016A}, as well as non-fermi liquid in  itinerant quantum critical regions~\cite{Wang2016,Lederer2016,XuXiaoYan2017b}. The strong-coupling nature of these systems makes analytical approach challenging~\cite{SungSik2009,Sachdev2010A,Sachdev2010B,Dalidovich2013,Schlief2016}, and hence sign-problem-free QMC solutions pave a new avenue towards quantitative understanding about these systems. These QMC approaches also offer a new platform for studying strongly-correlated topological states, and have recently been utilized to study topological phase transitions in DSM~\cite{XuXiaoYan2017a} and exotic states with topological order~\cite{Assaad2016,Sachdev2016,Gazit2017}.

\begin{figure*}
\centering
\includegraphics[width=0.8\textwidth]{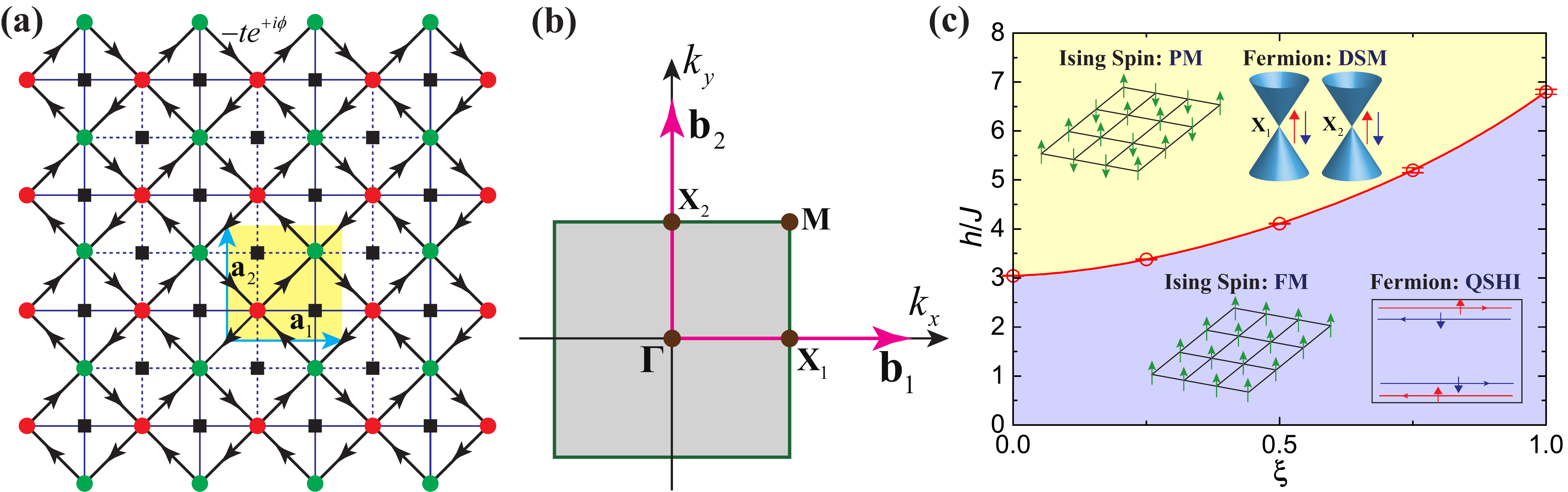}
\caption{\label{fig:LatticeBZPhaseDiagram} The checkerboard lattice and the ground-state phase diagram. (a) The checkerboard lattice (disks) and the dual
lattice (squares). Fermionic fields (Ising spins) are introduced to the lattice (dual lattice) sites. A unit cell of the lattice, indicated by $\mathbf{a}_1=(1,0),\mathbf{a}_2=(0,1)$, contains two fermion sites ($A$ and $B$ sublattices represented by red and green disks) and two Ising spins (black squares).
(b) The Brillouin zone. For $H_{\text{Fermion}}$, the band structure contains two Dirac points at
$\mathbf{X}_1$ and $\mathbf{X}_2$. (c) The phase diagram. Paramagnetic (PM) and ferromagnetic (FM)
phases of the Ising spins are separated by a continuous phase transition, which belongs to $N=8$ chiral Ising universality class with critical exponents $\nu=0.85(2)$ and $\eta=0.61(7)$ for finite $\xi$. At $\xi=0$, the phase transition belongs to 3D Ising universality class. In the
PM (FM) phase, fermions form a Dirac semimetal (quantum spin Hall insulator)}.
\end{figure*}

In this Letter, we study interaction-driven topological Mott insulators in Dirac semimetals with the aforementioned QMC approach. Instead of bare interactions, our model utilizes fluctuating bosonic fields to mediate interactions between fermions. At the level of the effective field theory, the model is equivalent to the originally proposed TMI model in Ref.~\cite{Raghu2008}, except for a minor difference in symmetry irrelevant to topology. For the study of TMI, our modified model shows two advantages: (1) other competing orders are strongly suppressed, allowing a clear TMI phase; (2) the sign-problem is avoided and thus the model can be solved via QMC techniques. Comparing to previous exact diagonalization studies~\cite{Valenzuela2013,Daghofer2014,Huaiming2014,Capponi2015,HanQing2016B}, the QMC approach can access much larger system size and reveals detailed information about the critical properties associated with the interaction-driven topological transition. Our QMC results show a continuous quantum phase transition from a DSM state to a QSH-type TMI phase, with the critical scaling at the quantum critical point agreeing nicely with the $N=8$ chiral Ising universality~\cite{Chandrasekharan2013,Otsuka2016}.

\textit{Model and Method}.
Our model describes Dirac fermions coupled to a transverse field Ising model. As illustrated in Fig.~\ref{fig:LatticeBZPhaseDiagram}(a), fermions in this model reside on the lattice sites (disks), while Ising spins are placed on each dual lattice site (squares) at the plaquette centers. The Hamiltonian consists of three parts,
\begin{eqnarray}
\label{eq:ModelHamiltonianMain}
H &=& H_{\text{Fermion}} + H_{\text{Ising}} + H_{\text{Coupling}},  \\ \nonumber
H_{\text{Fermion}} &=& -t\sum_{\langle ij \rangle\sigma} ( e^{+i\sigma\phi}c_{i\sigma}^{\dagger}c_{j\sigma} + e^{-i\sigma\phi}c_{j\sigma}^{\dagger}c_{i\sigma} ),    \\ \nonumber
H_{\text{Ising}} &=& -J\sum_{\langle pq\rangle}s_p^z s_q^z - h\sum_{p}s_p^x,   \\ \nonumber
H_{\text{Coupling}} &=& \sum_{\langle\langle ij \rangle\rangle\sigma}\xi_{ij} s_p^z( c_{i\sigma}^{\dagger}c_{j\sigma} + c_{j\sigma}^{\dagger}c_{i\sigma} ).
\end{eqnarray}
where indices $i,j$ represent fermion sites and $p,q$ label the dual lattice sites for Ising spins $s^z$. Fermion spins are labeled by subindex $\sigma$. $H_{\text{Fermion}}$ describes the nearest-neighbor (NN) hopping for fermions, which contains a staggered flux $\pm 4\phi$ for each plaquette. Here, we request spin-up and spin-down fermions to carrier opposite flux patterns to preserve the time-reversal symmetry. The Ising spins are governed by $H_{\text{Ising}}$, which describes a ferromagnetic ($J>0$) transverse-field Ising model. The last term $H_{\text{Coupling}}$ couples the Ising spins with the next-nearest-neighbor (NNN) fermion hoppings, where the coupling constant $\xi_{ij}=\pm \xi t$ has a staggered sign structure alternating between neighboring plaquette, i.e., $+$ ($-$) for solid (dashed) NNN bonds as illustrated in Fig.~\ref{fig:LatticeBZPhaseDiagram}(a). Up to a basis change, the low-energy physics in this model can be described by the following effective field theory $S=\sum_\sigma \int \mathbf{dr} dt  \bar{\Psi}_\sigma  (i \gamma^{\mu} \partial_\mu + g \sigma \varphi \gamma^{3}\gamma^{5} ) \Psi_\sigma+S_{\varphi}$, where $\gamma^\mu$ are gamma matrices and $\varphi$ is a bosonic field governed by the $\varphi^4$-theory $S_{\varphi}$. Here, $\sigma=\pm 1$ (up or down) is the fermion spin index, and $g$ is the coupling constant for the boson-fermion interactions.
This effective field theory is in strong analogy to the model proposed early on in Ref.~\cite{Raghu2008}, provided that we decouple the fermion-fermion interactions with a Hubbard-Stratonovich auxiliary field,  as appropriate  in the limit $h/J \to\infty$~\cite{Assaad2016}. It is also worthwhile to emphasize that in our model, the fermion spins only preserve U(1) symmetry, while the model in Ref.~\cite{Raghu2008} has a SU(2) spin symmetry. This difference has little effect on topological properties, but as discussed below it changes the critical scaling as well as the finite temperature phase diagram.

As in the original model of TMI, our Hamiltonian also contains a symmetry which prohibits nontrivial topology. It is easy to verify that our Hamiltonian is invariant under the following $Z_2$ transformation, $\hat{P}=\hat{R}_x(\pi)\times \hat{T}_{A\to B}$, where $\hat{R}_x(\pi)$ stands for $\pi$-rotation along $x$-axis for both Ising and fermion spins, and $\hat{T}_{A\to B}$ represents space translation from sublattice $A$ to $B$ inside a unit cell. Because the topological index (the spin Chern number) flips sign under this transformation, this symmetry requires the index to vanish and thus any (quantum spin Hall) topological insulator is prohibited, unless this $Z_2$ symmetry is broken spontaneously.

To explore the ground-state phase diagram of this model, we employ the projector quantum Monte Carlo (PQMC) method~\cite{AssaadEvertz2008}, with details presented in Sec.I.A of the supplemental material (SM)~\cite{Suppl}. In addition to the usual local updates of Ising spins, both Wolff~\cite{Wolff1989} and geometric cluster updates~\cite{Heringa1998} are applied in our simulations, as shown in Sec.I.B of SM~\cite{Suppl}. Our QMC simulations are free of the sign problem at and away from half filling~\cite{Wucj2005}. In this Letter, we focus on the coupling strength $0\le\xi\le 1$ with $J=t=1$ and the system sizes simulated in this work are $L=4,6,8,10,12,14$ with $N=L^2$ unit cells and $N_s=2L^2$ lattice sites.

\textit{Ground state phase diagram}. The ground state phase diagram in the  $\xi-h$ plane  is shown in Fig.~\ref{fig:LatticeBZPhaseDiagram}(c). Several regimes in the phase diagram can be solved exactly. At $\xi=0$, the fermions and Ising spins decouple: the fermions form a non-interacting Dirac semimetal, and the  Ising spins undergo a paramagnetic to ferromagnetic (PM-FM) quantum phase transition at $h_c=3.046(3)$  in the  3D Ising universality class~\cite{Pfeuty1971,XuXiaoYan2017a}. At $h=0$, quantum fluctuations of Ising spins vanish and Ising spins form a fully-polarized FM state. As a result, the fermions turn into a non-interacting quantum-spin-Hall topological insulator, whose Hamiltonian is $H_{\text{Fermion}}+H_{\text{Coupling}}$ with fully polarized Ising spins $s^z=+1(-1)$ ~\cite{Sun2009,Hou2013,HanQing2016B} (See Sec. V.A in the SM~\cite{Suppl} for details). At $h\to\infty$, the Ising spins are aligned along the $x$-axis. Second order perturbation theory around this point,  gives rise to an interaction of order $\xi^2/h$ between the fermions. Since the Dirac semimetal is a stable state of matter, we expect that it will be realized in the limit $h \rightarrow \infty$.

At $\xi>0$ and intermediate $h$, we find a direct second-order quantum transition between the PM and FM phases. This transition is also the topological phase transition for the fermions, in which the Dirac semimetal acquires a topological mass gap corresponding to the quantum spin Hall topological insulator. This conclusion is consistent with the symmetry analysis above, where the PM (FM) phase preserve (spontaneously breaks) the $Z_2$ symmetry and thus a quantum spin Hall insulator is prohibited (allowed). At $\xi>0$, the scaling exponents at the transition deviates from the 3D Ising universality class. Due to the coupling between fermions and bosons, the $\xi>0$ phase transition  flows to a different universality class, namely the $N=8$ component chiral Ising universality class~\cite{Chandrasekharan2013,Otsuka2016}.

\begin{figure}[t]
\centering
\includegraphics[width=0.8\columnwidth]{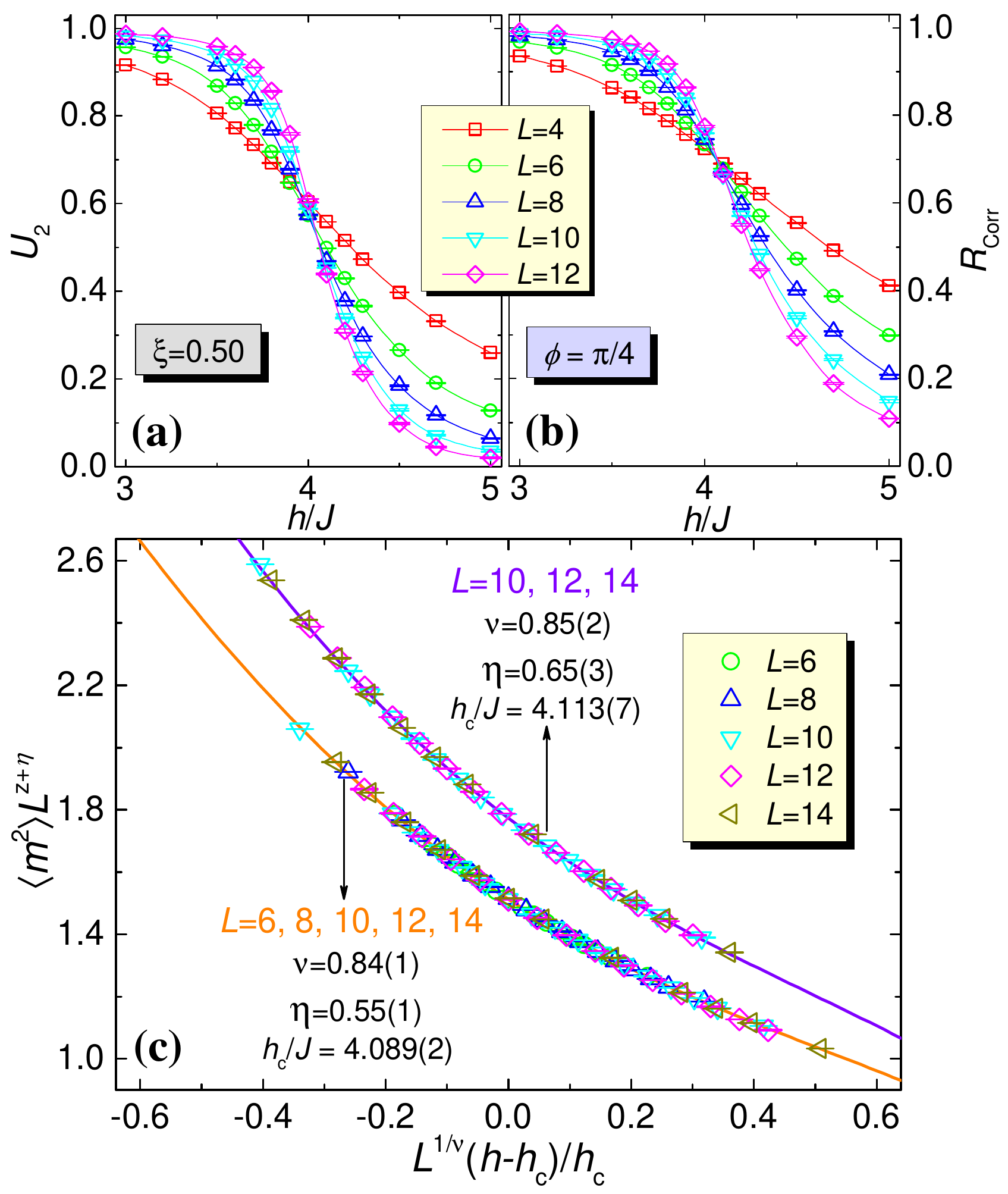}
\caption{\label{fig:DataCollapseXi050}(a) Binder cumulant $U_2$ and (b) correlation ratio $R_{\text{Corr}}$ for $\xi=0.50$ and $\phi=\pi/4$. (c) The data collapses of $\langle m^2 \rangle$ for $L=6,8,10,12,14$ and $L=10,12,14$, respectively. The critical exponents are also shown in (c).}
\end{figure}

\textit{FM-PM phase transition for Ising spins}. We determine the location of QCP via the Binder cumulant~\cite{Binder1981}: $U_2=\frac{1}{2}( 3 -  \frac{\langle m^4\rangle}{ \langle m^2 \rangle^2} )$ and correlation ratio~\cite{Kaul2015,Pujari2016}: $R_{\text{Corr}}=1-\frac{S^{\text{Ising}}(\mathbf{Q+\mathbf{q}})}{S^{\text{Ising}}(\mathbf{Q})}$, where $m=\frac{1}{N_s}\sum_{p}s_p^z$ and $S^{\text{Ising}}(\mathbf{k})$ is the trace of the structure factor matrix ($2\times 2$) of Ising magnetic order at $\mathbf{k}$ point. Here, $\mathbf{Q}=\boldsymbol{\Gamma}=(0,0)$ is the ordering vector for Ising spin, and $\mathbf{q}$ is the smallest momentum on the lattice, i.e., $(0,\frac{2\pi}{L})$ or $(\frac{2\pi}{L},0)$. Both $U_2$ and $R_{\text{Corr}}$ converge to 0 (1) in the PM (FM) phase at the thermodynamic limit. The crossing points for finite-size results of $U_2$ and $R_{\text{Corr}}$, respectively, provide the location of QCP. In this way, we first determine the position of QCP and then perform finite-size scaling analysis of $\langle m^2 \rangle$ close to it to extract the critical exponents.

The results of $U_2$ and $R_{\text{Corr}}$, as well as the data collapse of $\langle m^2 \rangle$ for $\xi=0.5$ and $\phi=\pi/4$ ($\pi$-flux in each plaquette) are presented in Fig.~\ref{fig:DataCollapseXi050}. Up to system size $L=12$, we can obtain the finite size crossing points $h=4.06$ for $U_2$ and $h=4.10$ for $R_{\text{Corr}}$ as the approximate location of QCP. In Fig.~\ref{fig:DataCollapseXi050}(c), we collapse the data as $\langle m^2\rangle L^{z+\eta}=f(L^{1/\nu}(h-h_c)/h_c)$ for $L=6,8,10,12,14$ and $L=10,12,14$, respectively. The critical exponents extracted from these two collapses are slightly different especially in $\eta$, indicating some finite-size effect. As will be discussed below, this shifting of exponents is due to a crossover phenomenon. Combining both collapses, we take the exponents as $\nu=0.85(2),\eta=0.61(7)$ (taking $z=1$) with $h_c=4.11(1)$, which are well consistent with the results presented in Ref.~\cite{Chandrasekharan2013} as $\nu=0.83(1),\eta=0.62(1)$ for $N=8$ components chiral Ising universality class.

We employed two additional measurements to further corroborate  the critical exponents. First, we performed finite-size scaling analysis for $S^{\text{Ising}}(\mathbf{k})$ at $\xi=0.50$ and $\phi=\pi/4$, which is shown in Sec. II.B of the SM~\cite{Suppl}, with the extracted critical exponents $\nu=0.84(4),\eta=0.62(6)$. Second, we also simulated the model with $\xi=0.50$ and $\phi=\pi/8$ (half-$\pi$ flux) and obtained the critical exponents from the finite-size scaling of $\langle m^2 \rangle$, and the results are presented in Sec. III.A of SM~\cite{Suppl}. The obtained critical exponents are $\nu=0.85(3),\eta=0.63(7)$ with $h_c=4.242(3)$. These exponents are well consistent with those in Fig.~\ref{fig:DataCollapseXi050}(c), rendering the $N=8$ components chiral Ising universality class.

The properties of QCPs for the PM-FM phase transitions of Ising spins for $\xi=0.25,0.75,1.00$ as presented in the phase diagram of Fig.~\ref{fig:LatticeBZPhaseDiagram}(c), are also determined with $U_2$ and $R_{\text{Corr}}$, as well as the finite-size scaling of $\langle m^2 \rangle$ and excitation gaps of fermions.

\begin{figure}[t]
\centering
\includegraphics[width=0.8\columnwidth]{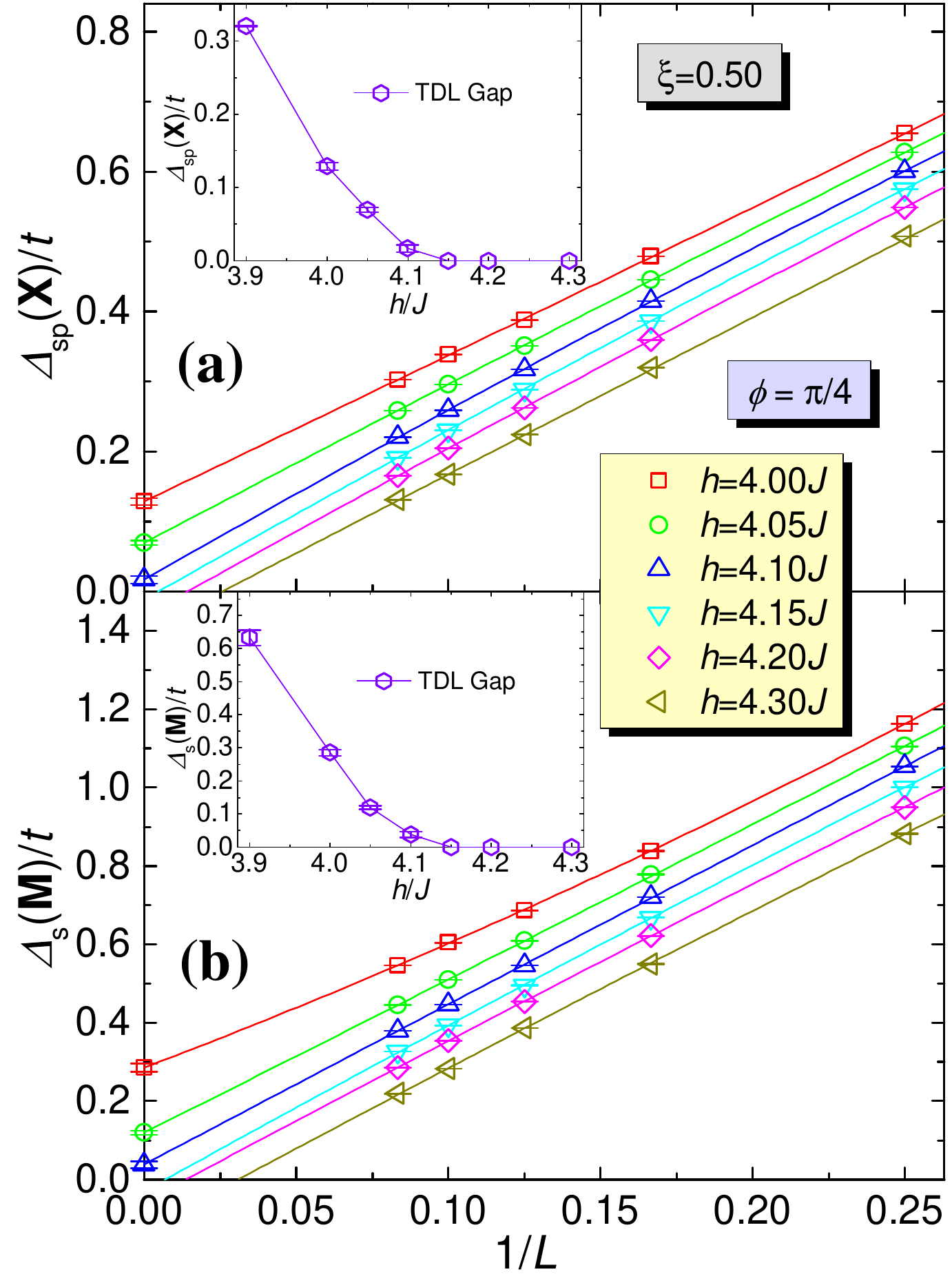}
\caption{\label{fig:ExcitationGapsXi050} (a) The single-particle gap $\Delta_{sp}(\mathbf{X})$ and (b) the spin gap $\Delta_s(\mathbf{M})$ close to the QCP $h_c=4.11(1)$ for $\xi=0.50$ and $\phi=\pi/4$. The insets are the excitation gaps at thermodynamic from the extrapolation with second-order polynomials in $1/L$. Both single-particle and spin gaps open at $h/J=4.10\sim4.15$, consistent with QCP of PM-FM phase transition of Ising spins.}
\end{figure}

\textit{Topological phase transition for fermions}. Our numerical results further show a single phase transition from Dirac semimetal to topological Mott insulator with decreasing transverse-field $h$, which comes hand in hand with the PM-FM phase transition of Ising spins. As shown in Fig.~\ref{fig:ExcitationGapsXi050}, we find that the fermions remains gapless in the PM phase with vanishing gap at the Dirac point. Here, both $\Delta_{sp}(\mathbf{X})$, the average single particle gap at two Dirac points $\mathbf{X}_1$ and $\mathbf{X}_2$, and $\Delta_s(\mathbf{M})$, the two-particle spin gap at $M$, vanish at the thermodynamic limit, consistent with the Dirac semimetal spectrum. In the FM phase, both gaps start to merge at $h/J=4.10\sim4.15$, consistent with the location of PM-FM phase transition point for Ising spins. It is worthwhile to highlight that the gaps remain finite in the whole FM phase with $h<h_c$, indicating the absence of  topological phase transition. Since the fermions form a quantum spin Hall insulator in the exactly solvable limit at $h=0$, this finite gap implies that the whole FM phase shares the same nontrivial topology. To further verify this conclusion, we compute directly the topological invariant, the spin Chern number $C_s=(C_{\uparrow}-C_{\downarrow})/2$. As shown in Sec. V.B. of SM~\cite{Suppl}, we obtain $C_s=+1$ for whole $h<h_c$ region, indicating that FM phase is a quantum-spin-Hall topological insulator.

In Sec. IV of SM~\cite{Suppl}, we present raw data of dynamic quantities $G(\mathbf{k},\tau)$ and $S^{xy}(\mathbf{k},\tau)$, from which $\Delta_{sp}(\mathbf{X})$ and $\Delta_s(\mathbf{M})$ are extrapolated. The comparisons between $2\Delta_{sp}(\mathbf{X})$ and $\Delta_s(\mathbf{M})$ are also shown to reveal the effect of electron-electron interactions. Furthermore, the gap opening of $\Delta_{sp}(\mathbf{X})$ and $\Delta_s(\mathbf{M})$ at $\xi=0.25,0.75,1.00$  match the QCPs of PM-FM phase transition for Ising spins, thus supporting  the picture of a semimetal-TMI topological phase transition.

\textit{Finite-size scaling crossover}. As discussed above at $\xi=0$ and $\xi>0$, the PM-FM transition belongs to two different universality classes, 3D Ising and $N=8$ chiral Ising. As a result, in the thermodynamic limit, the scaling exponents will change discontinuously as we change the value of $\xi$ away from $0$. In numerical studies, because of the finite size, such a discontinuous change will not show up. Instead, a crossover behavior is expected, i.e. at small $\xi$, a crossover length scale $L_c(\xi)$ shall arise. For $L<L_c$ ($L>L_c$), the scaling behavior merges towards the 3D Ising ($N=8$ chiral Ising) universality class. As $\xi$ approaches zero (increases), $L_c$ diverge to infinity (decreases to microscopic values) and thus the 3D Ising ($N=8$ chiral Ising) universality class is fully recovered. Such an effect is indeed observed in our data. In Sec. VI in SM~\cite{Suppl}, we present the finite-size scalings of $\langle m^2\rangle$ from $L=6,8,10,12$ and $L=8,10,12$, respectively, for $\xi=0.25,0.50,0.75$. At $\xi=0.25$, the data collapse suffers strongly from the finite-size effect, and chiral Ising exponents only arise in very large system sizes, especially for $\eta$. However, as $\xi$ increases, the chiral Ising exponents emerge even if the smallest size $L=6$ is included in the fitting.

\textit{Discussions}. Because the fermion spin in our model only preserves a $U(1)$ symmetry, instead of $SU(2)$, our topological Mott insulator breaks a $Z_2$ symmetry in contrast to the $SU(2)$ symmetry breaking in Ref.~\cite{Raghu2008}. This difference in symmetry breaking patterns is irrelevant for topology. However, this leads to different scaling exponents at the transition~\cite{Chandrasekharan2013}. Furthermore, at finite temperature, the symmetry breaking phase in our model survives, while the $SU(2)$ symmetry breaking arises only at $T=0$.

To the best of our knowledge, our study demonstrates the first interaction-driven quantum-spin-Hall topological Mott insulator from unbiased numerical method, and for the first time, this novel topological phenomenon becomes accessible to large-scale lattice QMC simulations. Our work points out  a new route to realize interaction-driven topological phases and phase transitions. It has experimental relevance since the interaction-driven quantum anomalous Hall effect  has recently being suggested in functionalized $\alpha$-Fe$_{2}$O$_{3}$ nanosheet~\cite{Liang2017}.

We (YYH, XYX, ZYM and ZYL) acknowledge fundings from the Ministry of Science and Technology of China through National Key Research and Development Program under Grant No. 2016YFA0300502 and from the National Science Foundation of China under Grant Nos. 91421304, 11421092, 11474356, 11574359, 11674370 as well as the National Thousand-Young Talents Program of China. Y.Y.H is also supported by the Outstanding Innovative Talents Cultivation Funded Programs 2016 of Renmin University of China. K.S. acknowledges support from the National Science Foundation under Grant No. PHY1402971 and the Alfred P. Sloan Foundation. F.F.A thanks the German Research Foundation (DFG) for financial support through  the SFB 1170 ToCoTronics. We thank the Physical Laboratory of High Performance Computing in Renmin University of China, the Center for Quantum Simulation Sciences in the Institute of Physics, Chinese Academy of Sciences and the Tianhe-1A platform at the National Supercomputer Center in Tianjin for their technical support and generous allocation of CPU time.

\bibliography{FermionIsingMain}


\onecolumngrid

\newpage

\begin{center}
\textbf{\large Supplemental Material: Chiral Ising transition between Dirac semimetal and quantum spin Hall insulators}
\end{center}

\author{Yuan-Yao He}
\address{Department of Physics, Renmin University of China, Beijing 100872, China}
\author{Xiao Yan Xu}
\address{Beijing National Laboratory for Condensed Matter Physics, and Institute of Physics, Chinese Academy of Sciences, Beijing 100190, China}
\author{Kai Sun}
\address{Department of Physics, University of Michigan, Ann Arbor, MI 48109, USA}
\author{Fakher F. Assaad}
\address{Institut f\"ur Theoretische Physik und Astrophysik, Universit\"at W\"urzburg, 97074 W\"urzburg, Germany}
\author{Zi Yang Meng}
\address{Institute of Physics, Chinese Academy of Sciences, Beijing 100190, China}
\affiliation{School of Physical Sciences, University of Chinese Academy of Sciences, Beijing, 100190, China}
\author{Zhong-Yi Lu}
\address{Department of Physics, Renmin University of China, Beijing 100872, China}

\setcounter{equation}{0}
\setcounter{figure}{0}
\setcounter{table}{0}
\setcounter{page}{1}
\makeatletter
\renewcommand{\theequation}{S\arabic{equation}}
\renewcommand{\thefigure}{S\arabic{figure}}

\section{I. Projector QMC Algorithm for the model}
\label{sec:PQMCIntroduction}

\subsection{A. PQMC Formalism}
\label{sec:PQMCFormalism}

\hspace{0.5cm} The fermion-Ising model in the main text is expressed as
\begin{eqnarray}
\label{eq:ModelHamiltonianSup}
H &=& H_{\text{Fermion}} + H_{\text{Ising}} + H_{\text{Coupling}},  \\ \nonumber
H_{\text{Fermion}} &=& -t\sum_{\langle ij \rangle\sigma} ( e^{+i\sigma\phi}c_{i\sigma}^{\dagger}c_{j\sigma} + e^{-i\sigma\phi}c_{j\sigma}^{\dagger}c_{i\sigma} ),    \\ \nonumber
H_{\text{Ising}} &=& -J\sum_{\langle pq\rangle}s_p^z s_q^z - h\sum_{p}s_p^x,   \\ \nonumber
H_{\text{Coupling}} &=& \sum_{\langle\langle ij \rangle\rangle\sigma} \xi_{ij}\xi s_p^z( c_{i\sigma}^{\dagger}c_{j\sigma} + c_{j\sigma}^{\dagger}c_{i\sigma} ),
\end{eqnarray}
In the projector QMC setup, the partition function of this model can be expressed as
\begin{eqnarray}
\label{eq:Partition00}
\mathcal{Z} = \langle\psi_T| Tr_{\{\mathbf{s}\}} \big\{e^{-\Theta H}\big\} |\psi_T\rangle
\end{eqnarray}
where the trace is taken over all Ising spin configurations, and $|\Psi_T \rangle$ is the Slater-determinant trial wave function.

\hspace{0.5cm} In projector QMC framework, we first break the projection length $\Theta$ into $M$ slices ($\Theta=M\Delta \tau$). Since the $s_p^z$ operator in $H_{\text{Coupling}}$ term is diagonal under Ising spin configurations, the Boltzmann weight from $H_{\text{Ising}}$ is separable. After tracing out fermion degrees of freedom, we can  obtain the partition function of the fermion-Ising coupling model as~\cite{XuXiaoYan2017a}
\begin{eqnarray}
\label{eq:Partition01}
\mathcal{Z} &=& \langle\psi_T| Tr_{\{\mathbf{s}\}} \big\{ e^{-\Theta\hat{H}} \big\} |\psi_T\rangle  \\ \nonumber
 &\approx& \sum_{\{s_{i,\ell}^z=\pm1\}} \ \mathcal{W}_{\text{Ising}} \prod_{\sigma=\uparrow,\downarrow}\det(P_{\sigma}^{\dagger}B_M^{\sigma}\cdots B_2^{\sigma}B_1^{\sigma}P_{\sigma}).
\end{eqnarray}
where
\begin{equation}
\mathcal{W}_{\text{Ising}}  =
\lambda^{MN_s} \exp\Big[-\Big( -\Delta\tau J  \sum_{\ell=1}^M\sum_{\langle pq \rangle} s_{p,\ell}^z s_{q,\ell}^z - \gamma\sum_p\sum_{\ell=1}^M s_{p,\ell+1}^z s_{p,\ell}^z \Big)\Big]
\end{equation}
is from $H_\text{Ising}$, when the 2D transverse-field Ising model is mapped to a 3D classical Ising model~\cite{Isakov2003,Youjin2002,WangYanCheng2016,XuXiaoYan2017a}, with
\begin{eqnarray}
\left\{\begin{array}{ll}
    \lambda = \sqrt{\sinh(\Delta\tau h) \cosh(\Delta\tau h)}  \\
    \gamma = -\frac{1}{2}\ln[\tanh(\Delta\tau h)]
\end{array}.
\right.
\end{eqnarray}
Due to the spin-staggered phase $e^{i\sigma\phi}$ in $H_{\text{Fermion}}$ term in Eq.~(\ref{eq:ModelHamiltonianSup}), the spin-up determinant $\det(P_{\uparrow}^{\dagger}B_M^{\uparrow}\cdots B_2^{\uparrow}B_1^{\uparrow}P_{\uparrow})$ in Eq.~(\ref{eq:Partition01}) is complex-conjugate to the spin-down one $\det(P_{\downarrow}^{\dagger}B_M^{\downarrow}\cdots B_2^{\downarrow}B_1^{\downarrow}P_{\downarrow})$. Since the configuration weight for the classical Ising spin part is always positive, the configuration weight containing both the fermion and classical spin parts is non-negative thus there is no sign problem in the QMC simulations of the model Hamiltonian in Eq.~(\ref{eq:ModelHamiltonianSup}). To calculate the $B_{\ell}^{\sigma}$, we have applied Trotter decomposition for both the $e^{-\Delta\tau H_{\text{Fermion}}}$ and $e^{-\Delta\tau H_{\text{Coupling}}}$ terms. For the $H_{\text{Fermion}}$ term, we have applied the checkerboard decomposition, which divides all the NN hopping terms into two parts such that in each part all the hopping terms commute with one another. For the $H_{\text{Coupling}}$ term as the fermion-Ising spin coupling term, it's separated into four parts such that in each part all the hopping term commute with one another. Including both of these decompositions, we have the $B_{\ell}^{\sigma}$ matrix as
\begin{eqnarray}
B_{\ell}^{\sigma} = e^{\mathcal{C}_{\ell,4}}e^{\mathcal{C}_{\ell,3}}e^{\mathcal{C}_{\ell,2}}e^{\mathcal{C}_{\ell,1}} e^{\mathcal{K}_2} e^{\mathcal{K}_1},
\end{eqnarray}
where $\mathcal{C}_{\ell,4},\mathcal{C}_{\ell,3},\mathcal{C}_{\ell,2},\mathcal{C}_{\ell,1}$ correspond to the $H_{\text{Coupling}}$ term while $\mathcal{K}_2,\mathcal{K}_1$ represent the $H_{\text{Fermion}}$ term. Here, all of $\mathcal{C}_{\ell,4},\mathcal{C}_{\ell,3},\mathcal{C}_{\ell,2},\mathcal{C}_{\ell,1}$ and $\mathcal{K}_2,\mathcal{K}_1$ are effectively $4\times4$ matrices.

\hspace{0.5cm} For the results presented in main text, we set $\Delta\tau=0.04/t$ and purposely increase the projection parameter $\Theta$ for increasing system size, such as $\Theta=20$ for $L=4$ and $\Theta=50$ for $L=14$, to ensure convergence to ground state.

\subsection{B. Cluster Updates for Ising spins}
\label{sec:ClusterUpdate}

\hspace{0.5cm} As for the updates of the classical Ising spins during the PQMC simulations, in one sweep, we first apply the local updates to scan through the space-time of configuration and then take one cluster update. For the case studied here, the 3D classical Ising model (2D transverse field Ising) has ferromagnetic interactions in both space and time directions as
\begin{eqnarray}
\beta^{\prime}\hat{H}^{\prime} = -\Delta\tau J\sum_{\ell=1}^M\sum_{\langle pq\rangle} s_{p,\ell}^z s_{q,\ell}^z - \gamma\sum_p\sum_{\ell=1}^M s_{p,\ell+1}^z s_{p,\ell}^z,
\end{eqnarray}
with $J>0$ and $\gamma=-\frac{1}{2}\ln[\tanh(\Delta\tau h)]>0$. This is a highly anisotropic model and the ratio between the coupling strengths in time and space is $r=\gamma/(\Delta\tau J)$, which can reach $10^1\sim10^3$. To guarantee the efficiency and correctness, we employ local updates, Wolff cluster updates~\cite{Wolff1989} and geometric cluster update~\cite{Heringa1998} in the simulation. The local update is, of course, the simplest one to be implemented. The two cluster updates, greatly suppress the critical slowing down, are complementary. As close to the quantum critical point, the size of Wolff cluster can sometimes be as large as the entire space-time lattice, rendering the update actually with low efficiency around the quantum criticality. The geometric cluster updates implicitly impose a restriction on the size of the cluster, making it not to grow too large. On the other hand, the geometric cluster updates can not change the magnetization of the system, but Wolff cluster update can explicitly change the value of $\sum_p s_{p,\ell}^z$.

\hspace{0.5cm} In PQMC, a sweep contains the propagation of the configurations in the imaginary time from $\tau=\Theta$ to $\tau=0$ and successive propagation from $\tau=0$ to $\tau=\Theta$. In the propagations of both $\tau=\Theta \to0$ and $\tau=0\to\Theta$, the local updates of classical Ising spin are carried out sequentially according to the space-time lattice. Then at $\tau=0$, the cluster updates are performed to flip spins in a cluster constructed by Wolff and geometric algorithms.

The detailed balance condition of Wolff or geometric updates is given by
\begin{eqnarray}
\pi(a)P(a\to b)\mathcal{A}(a\to b) = \pi(b)P(b\to a)\mathcal{A}(b\to a),
\end{eqnarray}
where $a,b$ are configurations of Ising spins, and $P(a\to b)$, $\mathcal{A}(a\to b)$ are the priori and acceptance probabilities. With $\pi(a)=\pi_{s}(a)\pi_{f}(a)$ (where $\pi_s(a)$ is the weight of spin part and $\pi_f(a)$ is the weight of fermion part) as the total configuration weight, both Wolff and geometric cluster algorithms carefully choose the cluster such that the priori probability ratio has the following property
\begin{eqnarray}
\frac{P(b\to a)}{P(a\to b)} = \frac{\pi_s(a)}{\pi_s(b)}.
\end{eqnarray}
Then we can obtain the acceptance probability as
\begin{eqnarray}
\mathcal{A}(a\to b) = \min\Big\{ \frac{P(b\to a)}{P(a\to b)} \frac{\pi(b)}{\pi(a)} , 1 \Big\}
 = \min\Big\{ \frac{\pi_f(b)}{\pi_f(a)}, 1 \Big\}.
\end{eqnarray}
\begin{table}[h!]
\caption{\label{tab:table1}Examples of the acceptance (1) or rejectance (0) for cluster updates (first column), number of spin to be flipped (second column), size of Wolff cluster (third column) and size of the geometric cluster (fourth column).}
\begin{tabular}{|p{2cm}<{\centering}|p{2cm}<{\centering}|p{2cm}<{\centering}|p{2cm}<{\centering}|}
\hline
\centering
Accept/Reject &  $n_{\text{flip}}$   &   $n_{\text{Wolff}}$  &  $n_{\text{geometric}}$  \\
\hline
 0     &   2400    &    2398   &      2     \\
\hline
 0     &    174    &     170   &      4     \\
\hline
 1     &    993    &      11   &    982    \\
\hline
 1     &    119    &       7   &    112     \\
\hline
 1     &     74    &      46   &     28     \\
\hline
 1     &   3436    &       2   &   3434   \\
\hline
 1     &    330    &       4   &    326   \\
\hline
 0     &   3421    &    3411   &     10    \\
\hline
 1     &   1203    &      23   &   1180    \\
\hline
 1     &     38    &      24   &     14   \\
\hline
\end{tabular}
\end{table}
One can see that the acceptance probability is simply the ratio of fermion part of weights for the classical Ising spin configurations. At $\tau=0$, arbitrary number of steps of Wolff and geometric cluster updates can be performed and the overall acceptance probability is
\begin{eqnarray}
\mathcal{A}(a_1\to a_n) = \mathcal{A}(a_1\to a_2)\mathcal{A}(a_2\to a_3)\cdots \mathcal{A}(a_{n-1}\to a_n)  = \min\Big\{ \frac{\pi_f(a_n)}{\pi_f(a_1)}, 1 \Big\}.
\end{eqnarray}
Hence the implementation of cluster updates in PQMC simulations goes as follows, first construct the cluster of Ising spins according to Wolff and geometric cluster algorithms, and then calculate the ratio of fermion part of weights to obtain the acceptance probability, such that the cluster update can be decided to be accepted or rejected.

From our simulations, we find that, in one sweep, performing one Wolff cluster update and one geometric cluster update reaches the highest acceptance ratio for the overall cluster updates. For example, the acceptance ratio for cluster updates is typically four times larger than that for the local updates. Table~\ref{tab:table1} lists the number of spins flipped, size of the Wolff and geometric clusters, for linear system size $L=12$ with $\xi/t=0.50,h/J=3.35$ close to FM-PM phase transition. One can see that in one step of cluster update, thousands of Ising spins are flipped.

\section{II. Quantum Critical Points of Ising spins}
\label{sec:DetermQCPs}

In this section, we demonstrate some additional results for the FM-PM phase transition of the Ising spins. These include  Sec. II. A., the determination of the QCPs for $\xi=0.25,0.75,1.00$, via the Binder cumulant $U_2$ and correlation ratio $R_{\text{Corr}}$. Sec. II. B, data collapse of structure factor $S^{\text{Ising}}(\boldsymbol{\Gamma})/L^2$, to extract the critical exponents.

\subsection{A. QCPs for different $\xi$}
\label{sec:QCPsXis}

As shown in the main text, we first determine the FM-PM phase transition points from the crossing of the Binder cumulant $U_2$ and correlation ratio $R_{\text{Corr}}$, defined as
\begin{eqnarray}
\label{eq:Ratios}
U_2 &=& \frac{1}{2}\Big( 3 -  \frac{\langle m^4 \rangle}{\langle m^2 \rangle^2} \Big)  \\ \nonumber
R_{Corr} &=& 1-\frac{S^{\text{Ising}}(\mathbf{Q+\mathbf{q}})}{S^{\text{Ising}}(\mathbf{Q})},
\end{eqnarray}
with the Ising spin magnetization $m=\frac{1}{N_s}\sum_{p}s_p^z$ (here we have $N_s=2L^2$). And the structure factor of classical Ising spin is defined on checkerboard lattice (two sublattices) as
\begin{eqnarray}
\label{eq:StructFact}
S^{\text{Ising}}(\mathbf{k}) = \frac{1}{2N}\sum_{a=A,B}\sum_{mn} e^{-i\mathbf{k}\cdot(\mathbf{R}_m-\mathbf{R}_n)} \langle s_{ma}^z s_{na}^z \rangle,
\end{eqnarray}
\begin{figure}[h!]
\centering
\includegraphics[width=0.43\columnwidth]{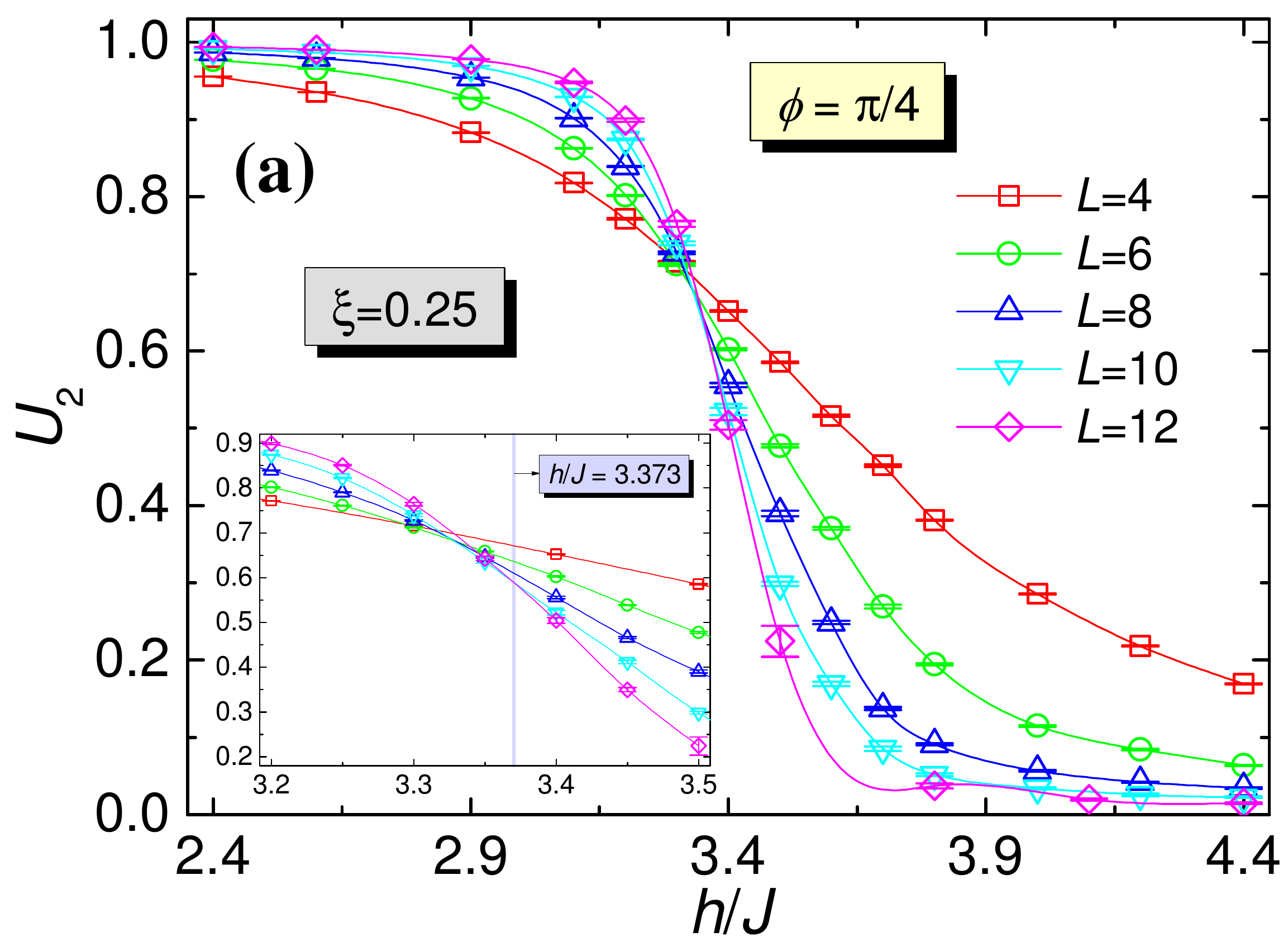}
\hspace{0.5cm}
\includegraphics[width=0.43\columnwidth]{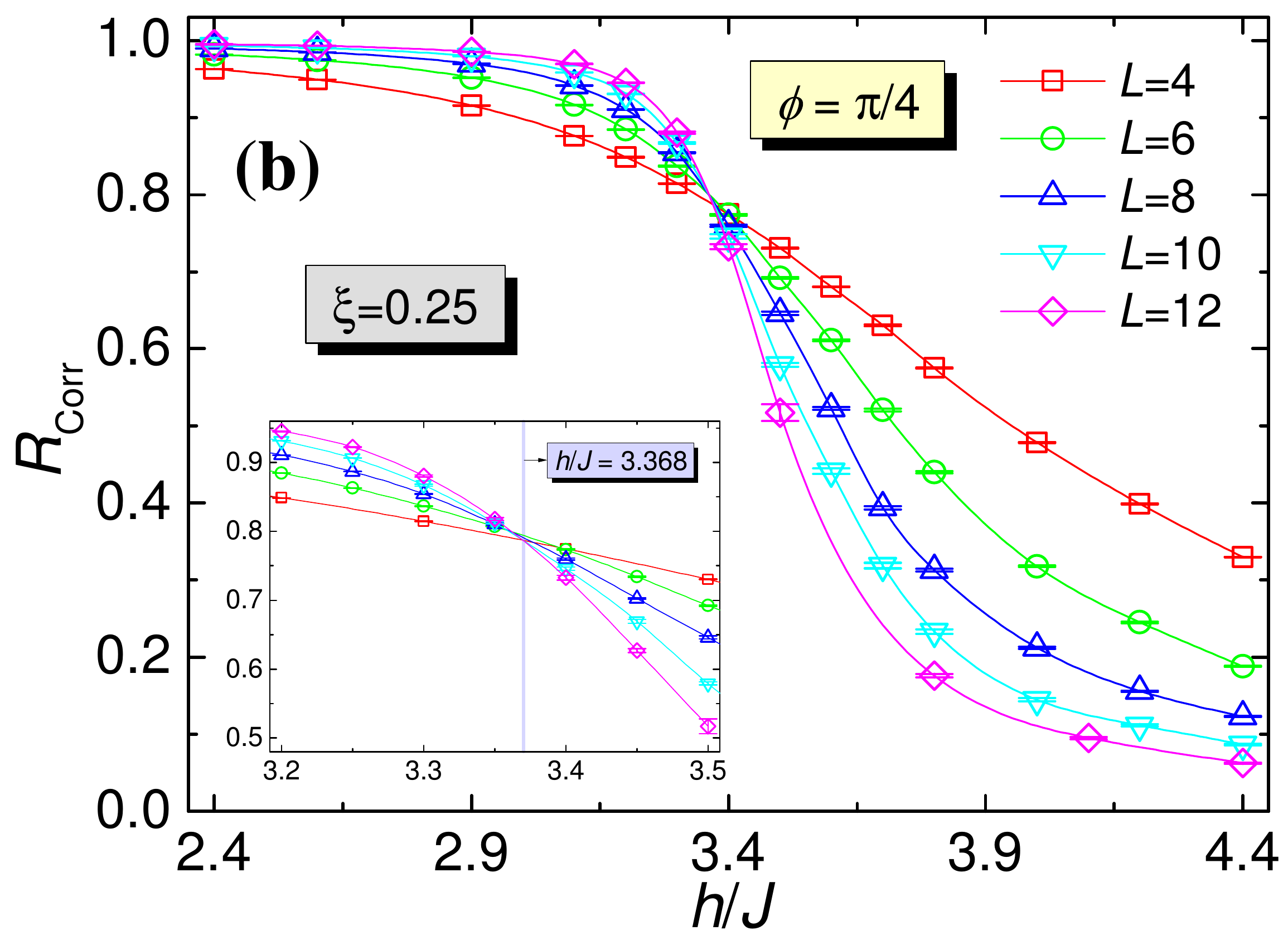}
\caption{\label{fig:RatioXi025} (a) Binder cumulant $U_2$ and (b) correlation ratio $R_{\text{Corr}}$ for $\xi=0.25$ and $\phi=\pi/4$ case across the QCP. }
\end{figure}
\begin{figure}[h!]
\centering
\includegraphics[width=0.43\columnwidth]{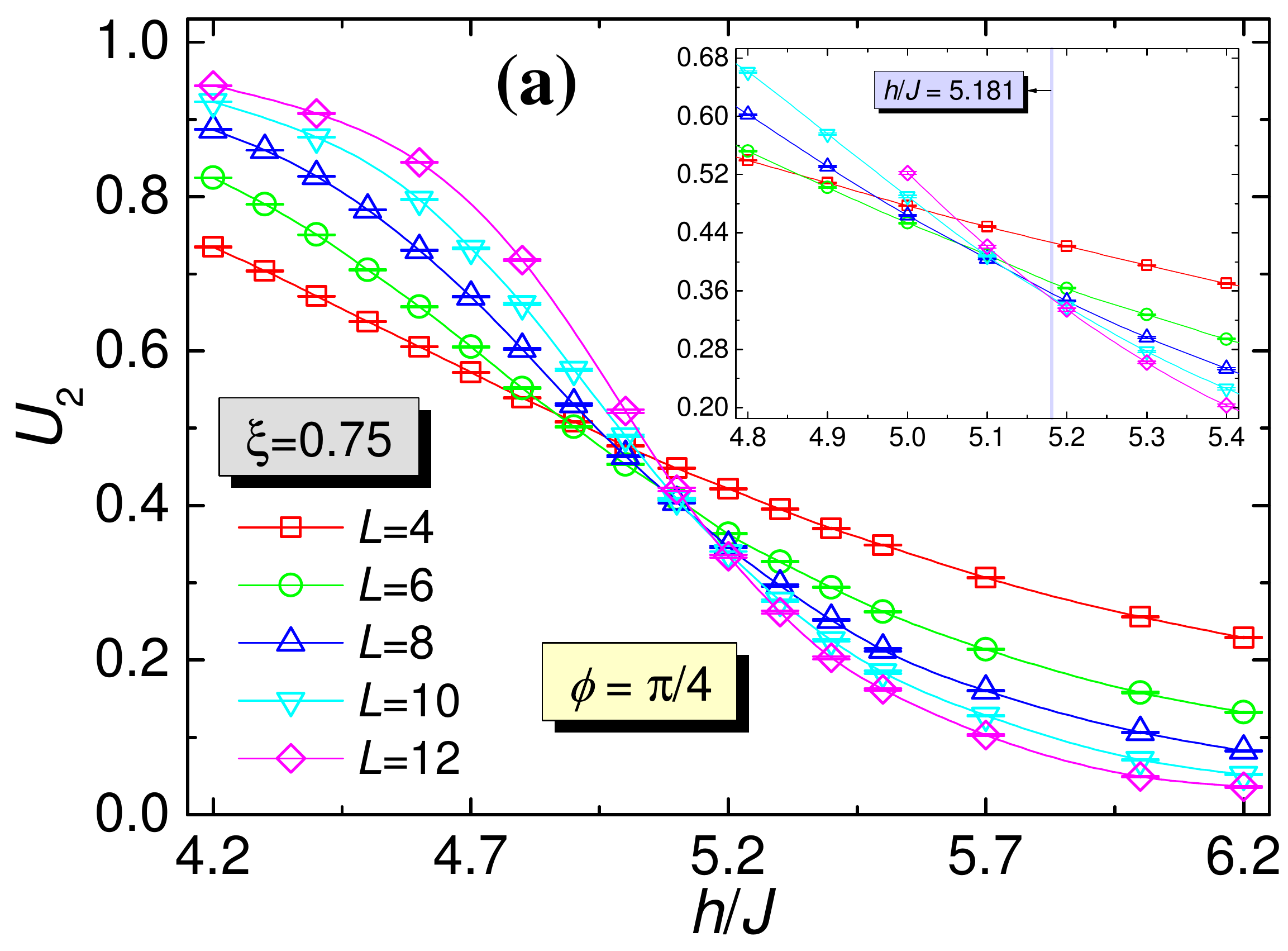}
\hspace{0.5cm}
\includegraphics[width=0.43\columnwidth]{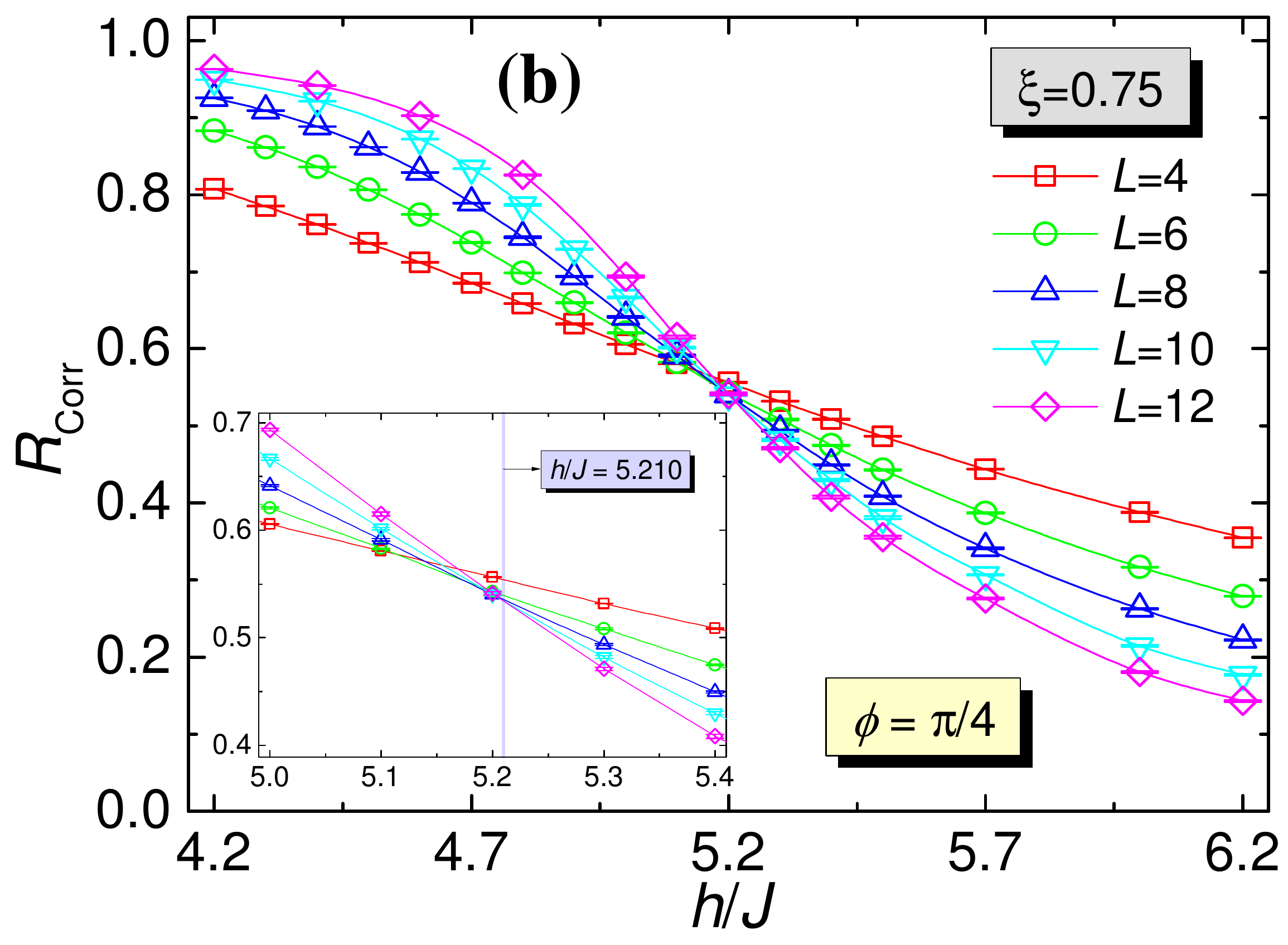}
\caption{\label{fig:RatioXi075} (a) Binder cumulant $U_2$ and (b) correlation ratio $R_{\text{Corr}}$ for $\xi=0.75$ and $\phi=\pi/4$ case across the QCP. }
\end{figure}
\begin{figure}[h!]
\centering
\includegraphics[width=0.43\columnwidth]{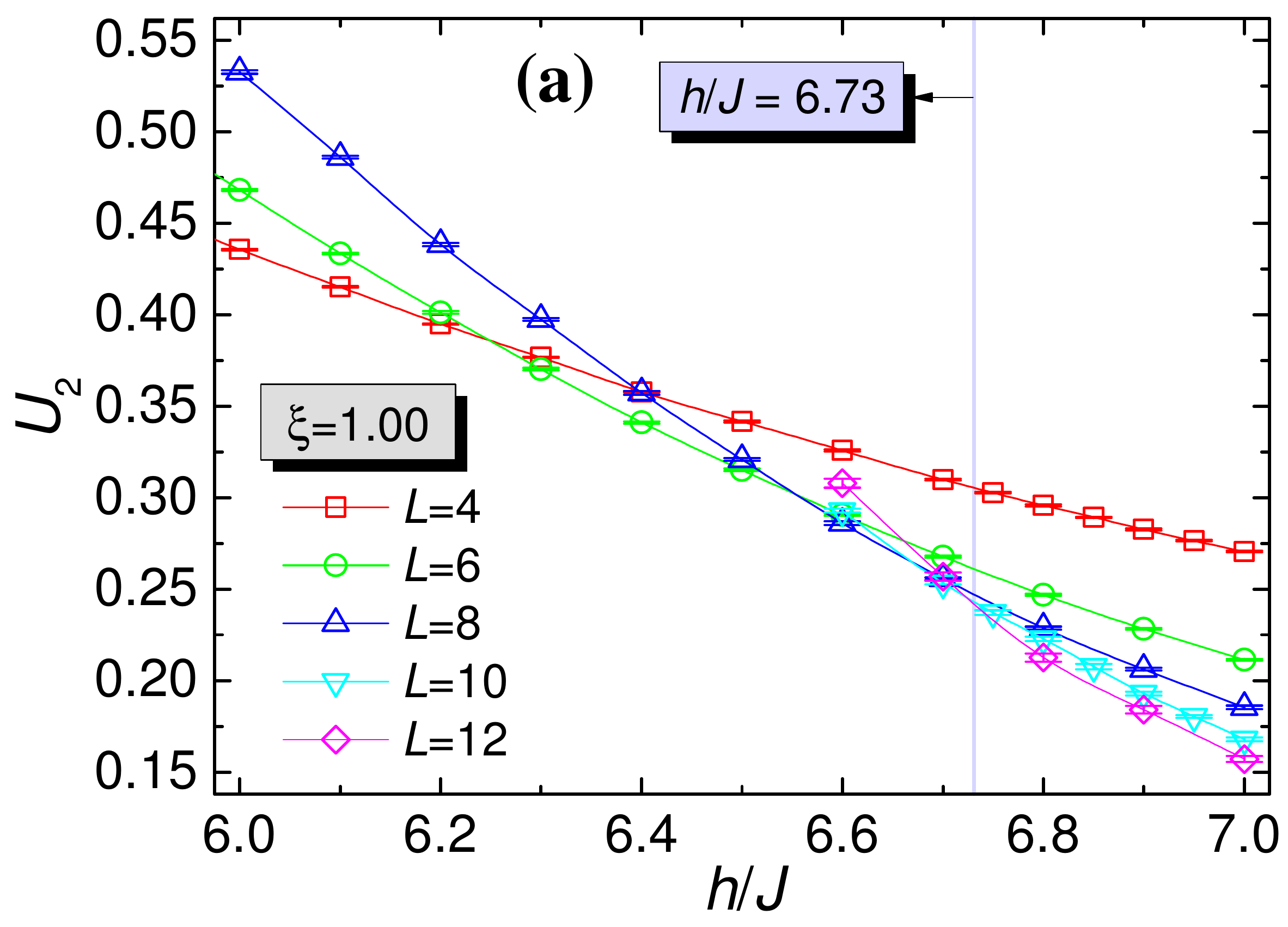}
\hspace{0.5cm}
\includegraphics[width=0.43\columnwidth]{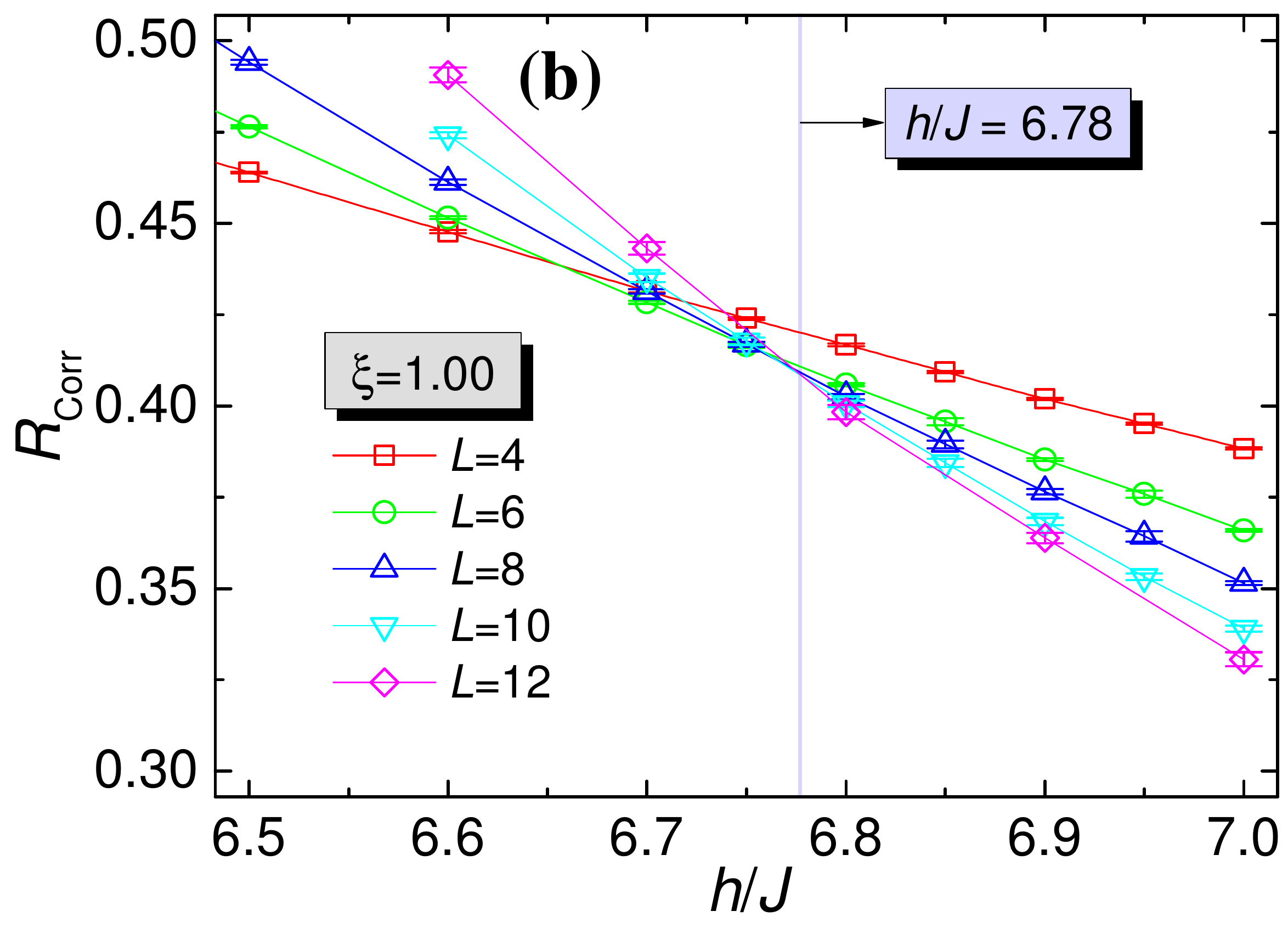}
\caption{\label{fig:RatioXi100} (a) Binder cumulant $U_2$ and (b) correlation ratio $R_{\text{Corr}}$ for $\xi=1.00$ and $\phi=\pi/4$ case across the QCP. }
\end{figure}
with $m,n$ as unit cell indexes and $a=A,B$ indicating the sublattice ($N=L^2$ as number of unit cells). In Eq.~(\ref{eq:Ratios}), $\mathbf{Q}$ is the ordered wave vector, which is $(0,0)$ for the ferromagnetic order, and $\mathbf{q}$ is the smallest distance vector away from $\mathbf{Q}$, equal to $(0,2\pi/L)$ or $(2\pi/L,0)$. It is expected that both $U_2$ and $R_{Corr}$ converge to 1 inside the ordered phase in the thermodynamic limit, while they are both 0 in disordered phase. So there is a jump from 1 to 0 when going from ordered phase to the disordered phase in thermodynamic limit, and for finite-size system they both show crossing between data with different system sizes, and the location of the crossing point defines the quantum critical point. Also, both $\langle m^2\rangle$ and $S^{\text{Ising}}(\boldsymbol{\Gamma})/L^2$ can tell whether the FM-PM phase transition is continuous or of first-order, depending on whether they are smooth or have drops close to the phase transition point.

In Fig.~\ref{fig:RatioXi025}, ~\ref{fig:RatioXi075} and ~\ref{fig:RatioXi100}, we present the results of both $U_2$ and $R_{\text{Corr}}$ for $\xi=0.25,0.75,1.00$ across the QCPs with $L=4,6,8,10,12$. Similar to the results of $\xi=0.50$, shown in the main text, both $U_2$ and $R_{\text{Corr}}$ converges to $1$ deeply inside the ordered phase and $0$ inside the disordered phase. The crossing points, for example $h/J=3.368,4.103,5.210,6.78$ for $L=10,12$ of $R_{\text{Corr}}$ for $\xi=0.25,0.50,0.75,1.00$, are good estimation of the QCPs, from which, we can further determine the more precise position of the thermodynamic QCPs by combining the data collapse of $\langle m^2 \rangle$ and extrapolation of excitation gaps.

\subsection{B. Data Collapse for Structure factor $S^{\text{Ising}}(\boldsymbol{\Gamma})/L^2$}
\label{sec:StructFactCollapse}

In the main text, we have obtained the critical exponents of the FM-PM phase transition of Ising spins from the data collapse of $\langle m^2\rangle$ data with the relation $\langle m^2\rangle L^{z+\eta}=f(\frac{h-h_c}{h_c}L^{1/\nu})$. In this part, we want to show that we can obtain the same results for $\nu,\eta$ exponents from the data collapse of $S^{\text{Ising}}(\boldsymbol{\Gamma})/L^2$ with similar relation as $L^{z+\eta}\cdot[S^{\text{Ising}}(\boldsymbol{\Gamma})/L^2]=f(\frac{h-h_c}{h_c}L^{1/\nu})$.
\begin{figure}[h!]
\centering
\includegraphics[width=0.64\columnwidth]{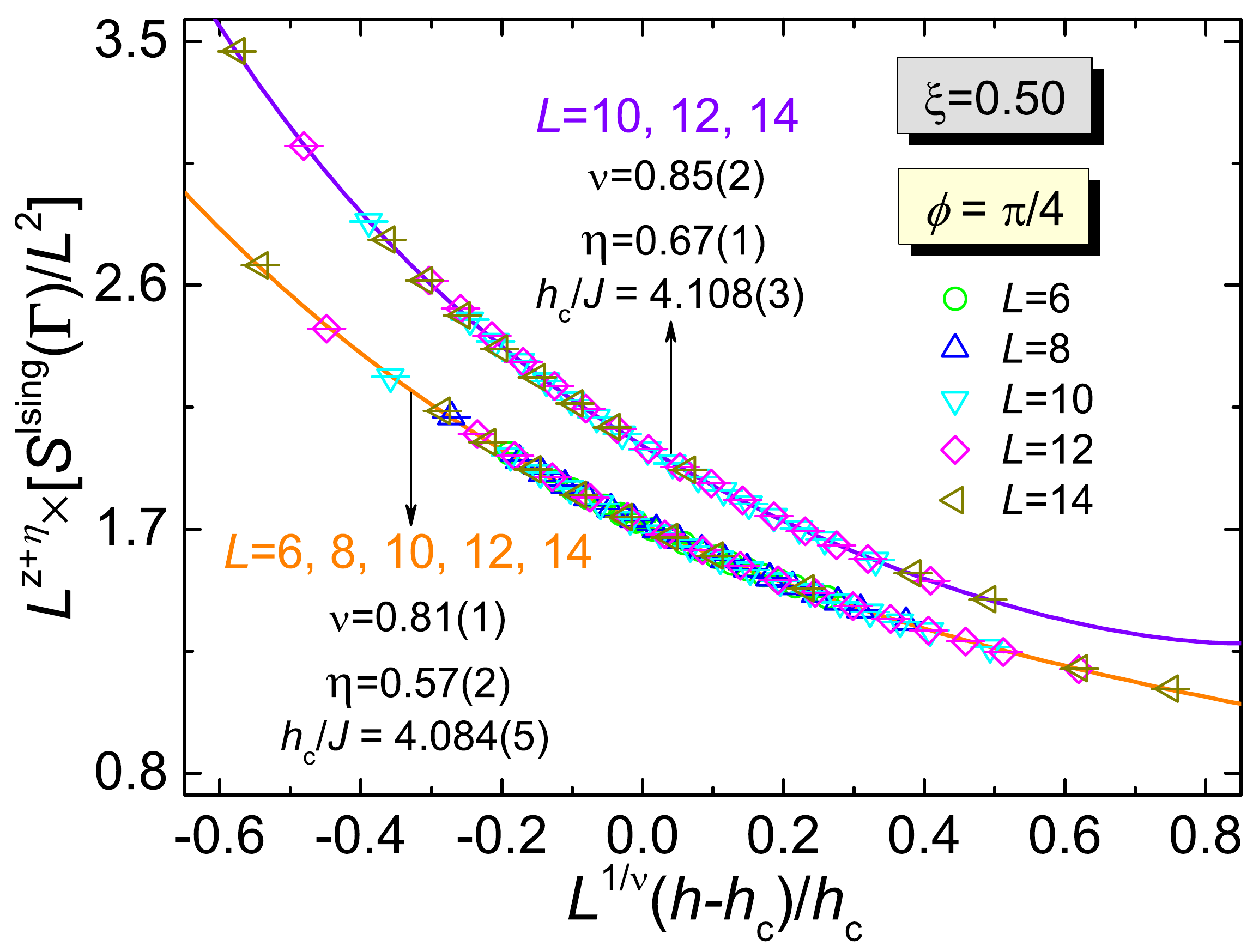}
\caption{\label{fig:StructFactCollapse} Data collapse of $S^{\text{Ising}}(\boldsymbol{\Gamma})/L^2$ for the QCP of FM-PM phase transition of Ising spins, with $\xi=0.50$ under $\pi$-flux case ($\phi=\pi/4$), for systems $L=6,8,10,12,14$. The critical exponents combining the collapses of data of $L=6,8,10,12,14$ and $L=10,12,14$ give rise to $\nu=0.84(4),\eta=0.62(6)$, which is well consistent with the numbers extracted from $\langle m^2\rangle$ data, as shown in the main text.}
\end{figure}
According to the definition of $S^{\text{Ising}}(\mathbf{k})$ in Eq.~(\ref{eq:StructFact}), we have
\begin{eqnarray}
\label{eq:StructFactGamma}
S^{\text{Ising}}(\boldsymbol{\Gamma})/L^2 = \frac{1}{2N^2}\sum_{a=A,B}\sum_{mn}  \langle s_{ma}^z s_{na}^z \rangle = \frac{1}{2}\Big\langle \Big(\frac{1}{N}\sum_{p\in A}s_p^z \Big)^2 + (\frac{1}{N}\sum_{p\in B}s_p^z \Big)^2 \Big\rangle = \frac{1}{2}\Big(\langle m_A^2\rangle + \langle m_B^2\rangle\Big),
\end{eqnarray}
with $m_{\alpha}=\frac{1}{N}\sum_{p\in\alpha}s_p^z$ ($\alpha=A,B$) sublattice magnetization for Ising spins. Similarly, we can express the $\langle m^2 \rangle$ by $m_A$ and $m_B$ as
\begin{eqnarray}
\label{eq:mSquareExpress}
\langle m^2 \rangle = \frac{1}{4}\Big( \langle m_A^2\rangle + \langle m_B^2\rangle + 2\langle m_A m_B\rangle \Big).
\end{eqnarray}
Then we can observe that at $h\to0$, the system is almost classically ordered and we have $\langle m_A^2\rangle=\langle m_B^2\rangle=\langle m_A m_B\rangle$ due to the classical decoupling of the correlation as $\langle m_A m_B\rangle=\langle m_A\rangle\langle m_B\rangle$. Thus, in the $h\to0$ limit, we have $\langle m^2 \rangle=S^{\text{Ising}}(\boldsymbol{\Gamma})/L^2$. However, for finite $h$, the equality doesn't hold anymore and $\langle m_A^2\rangle$ can be different from $\langle m_A m_B\rangle$ across the QCPs. However, they represent the same physical meaning of magnetization. Thus, we can surely perform the data collapse for them independently and the critical exponents extracted from them are expected to be the same.

Below we perform the data collapse for $S^{\text{Ising}}(\boldsymbol{\Gamma})/L^2$ for $\xi=0.50$ at $\pi$-flux case ($\phi=\pi/4$). The results are presented in Fig.~\ref{fig:StructFactCollapse}. The collapse of data from $L=6,8,10,12,14$ yields $\nu=0.81(1),\eta=0.57(2),h_c/J=4.084(5)$, while the collapse of data from $L=10,12,14$ gives $\nu=0.85(2),\eta=0.67(1),h_c/J=4.108(3)$ . Combining them, we conclude $\nu=0.84(4),\eta=0.62(6)$ for the FM-PM phase transition, which is well consistent with the results extracted from data collapse of $\langle m^2\rangle$, as shown in Fig.2 in main text.

\section{III. Quantum Phase Transitions for half-$\pi$ flux case ($\phi=\pi/8$)}
\label{sec:QPTHalfPi}

To further confirm the general properties of both Ising spins and fermions across the QCPs, we have also simulated the half-$\pi$-flux case as $\phi=\pi/8$. In Sec. III. A., we show that the FM-PM phase transition has almost the same critical exponents as that for the $\phi=\pi/4$ case, indicating the same universality class. What's more, the excitation gaps of the fermions also show gap opening behavior with decreasing $h$, suggesting the same DSM-TMI phase transition as that for $\phi=\pi/4$ case, which are presented in Sec. III. B. For this half-$\pi$ flux case ($\phi=\pi/8$), we only concentrate on $\xi=0.50$.

\subsection{A. FM-PM phase transition for Ising spins}
\label{sec:QPTHalfPiIsing}

\begin{figure}[h!]
\centering
\includegraphics[width=0.43\columnwidth]{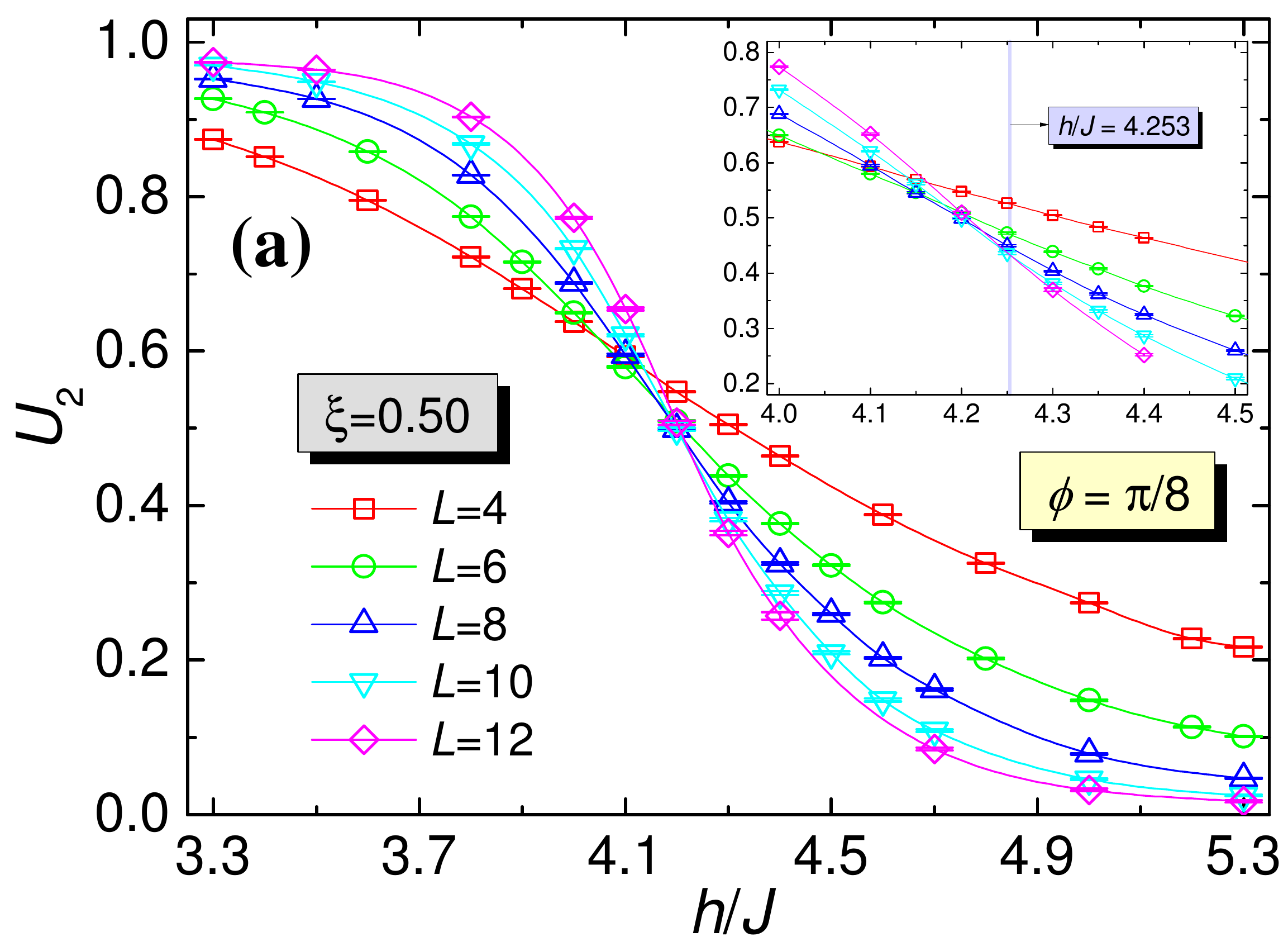}
\hspace{0.5cm}
\includegraphics[width=0.43\columnwidth]{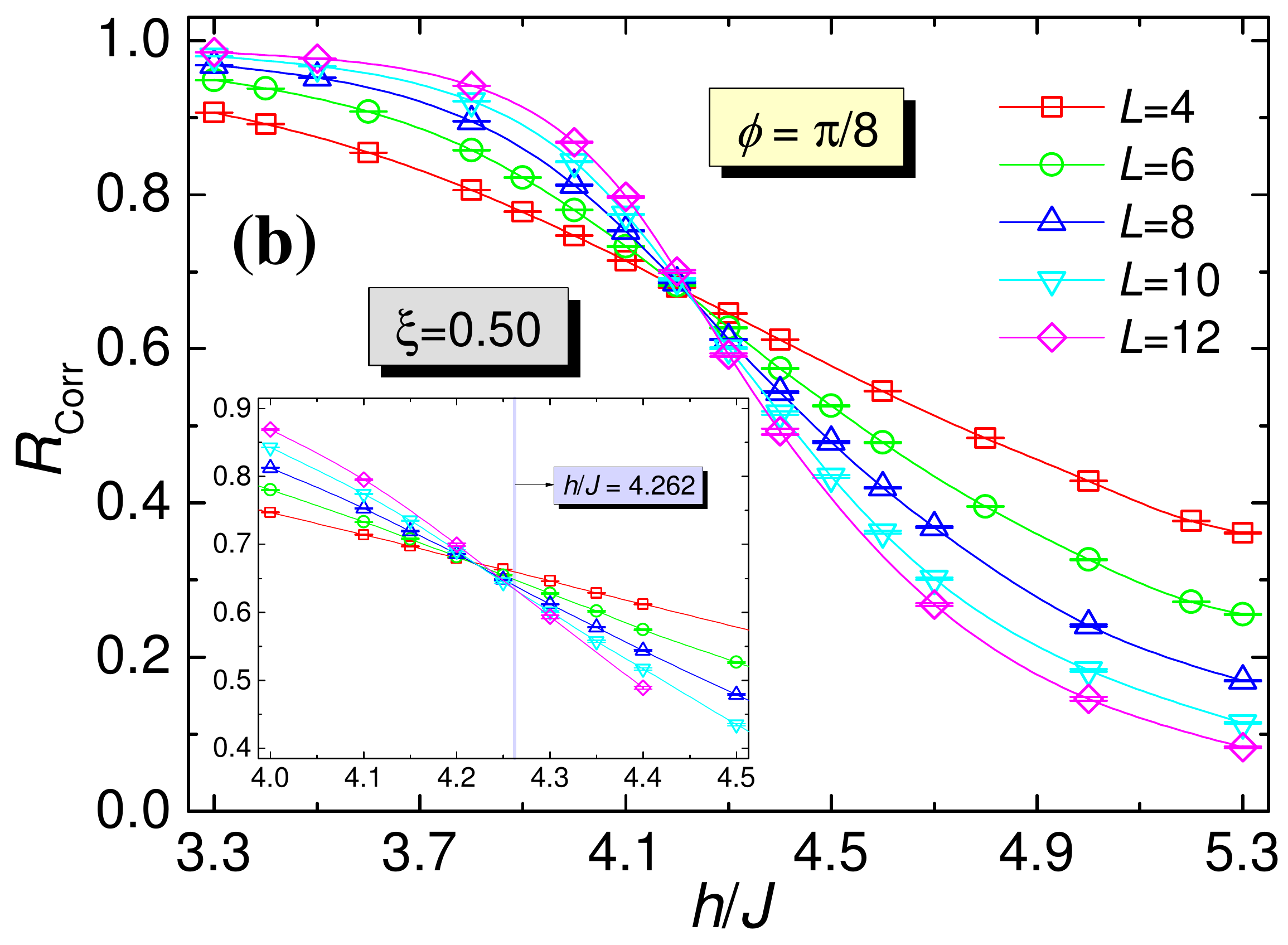}
\caption{\label{fig:RatioXi050HalfPi}(a) Binder cumulant $U_2$ and (b) correlation ratio $R_{\text{Corr}}$ for $\xi=0.50$ and $\phi=\pi/8$ across the QCP. }
\end{figure}

Again, the location of the QCP is determined from the crossings of $U_2$ and $R_{\text{Corr}}$ for different system sizes, which are shown in Fig.~\ref{fig:RatioXi050HalfPi}. Considering the convergence to thermodynamic limit, the QCP of this $\xi=0.50$ with $\phi=\pi/8$ should be $h_c/J>4.262$ (see the insets of Fig.~\ref{fig:RatioXi050HalfPi}), which is larger than that for $\phi=\pi/4$ case with the same $\xi$ parameter.

\begin{figure}[h!]
\centering
\includegraphics[width=0.54\columnwidth]{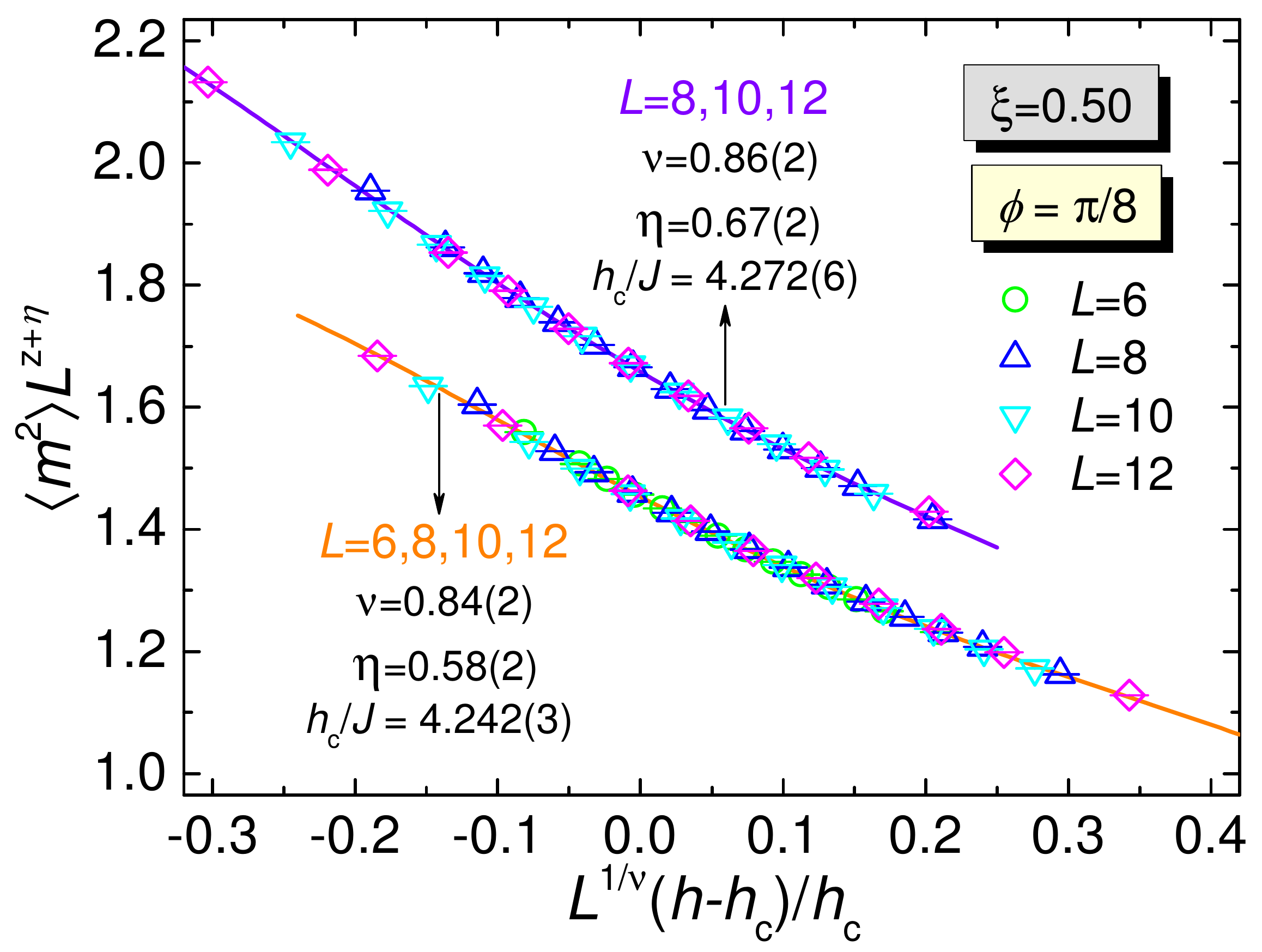}
\caption{\label{fig:Xi050HalfPi}Data collapse of $\langle m^2\rangle$ for the QCP of FM-PM phase transition for Ising spins, with $\xi=0.50$ for half-$\pi$-flux case ($\phi=\pi/8$), with $L=6,8,10,12$. The critical exponents thus obtained, combining the collapses of data of $L=6,8,10,12$ and $L=8,10,12$, are $\nu=0.85(3),\eta=0.63(7)$, well consistent with those extracted from $\phi=\pi/4$ case, as shown in the main text as well as in Fig.~\ref{fig:StructFactCollapse}.}
\end{figure}

Then we take a closer look at the QCP and perform the data collapse of $\langle m^2\rangle$ around the QCP. The results are shown in Fig.~\ref{fig:Xi050HalfPi}. Again, we have performed the collapse of the data for $L=6,8,10,12$ and $L=8,10,12$ to obtain the reasonable values of $\nu$ and $\eta$. The collapse of data from $L=6,8,10,12$ yields $\nu=0.84(2),\eta=0.58(2),h_c/J=4.242(3)$, while the collapse of data from $L=8,10,12$ gives $\nu=0.86(2),\eta=0.67(2),h_c/J=4.272(6)$. Combining them, we conclude $\nu=0.85(3),\eta=0.63(7)$ for the FM-PM phase transition, which is well consistent with the results extracted from data collapse of $\phi=\pi/4$ case, as shown in Fig.2 in main text as well as Fig.~\ref{fig:StructFactCollapse} in Sec. II. B. These results imply that both the FM-PM phase transitions of Ising spins for $\phi=\pi/4$ and $\phi=\pi/8$ cases belong to the $N=8$ Chiral Ising universality class, with  critical exponents consistent with previous work in Ref.~\onlinecite{Chandrasekharan2013,Otsuka2016}.

\subsection{B. DSM-TMI phase transition for fermions}
\label{sec:QPTHalfPiFermions}

Besides the PM-FM phase transition for the Ising spins, we have also confirmed it's accompanied by the DSM-TMI phase transition, via monitoring the opening of excitation gaps. We have calculated the single-particle gap $\Delta_{sp}(\mathbf{X})$ (average over $\mathbf{X}_1$ and $\mathbf{X}_2$ points) and spin gap $\Delta_s(\mathbf{M})$.
\begin{figure}[h!]
\centering
\includegraphics[width=0.43\columnwidth]{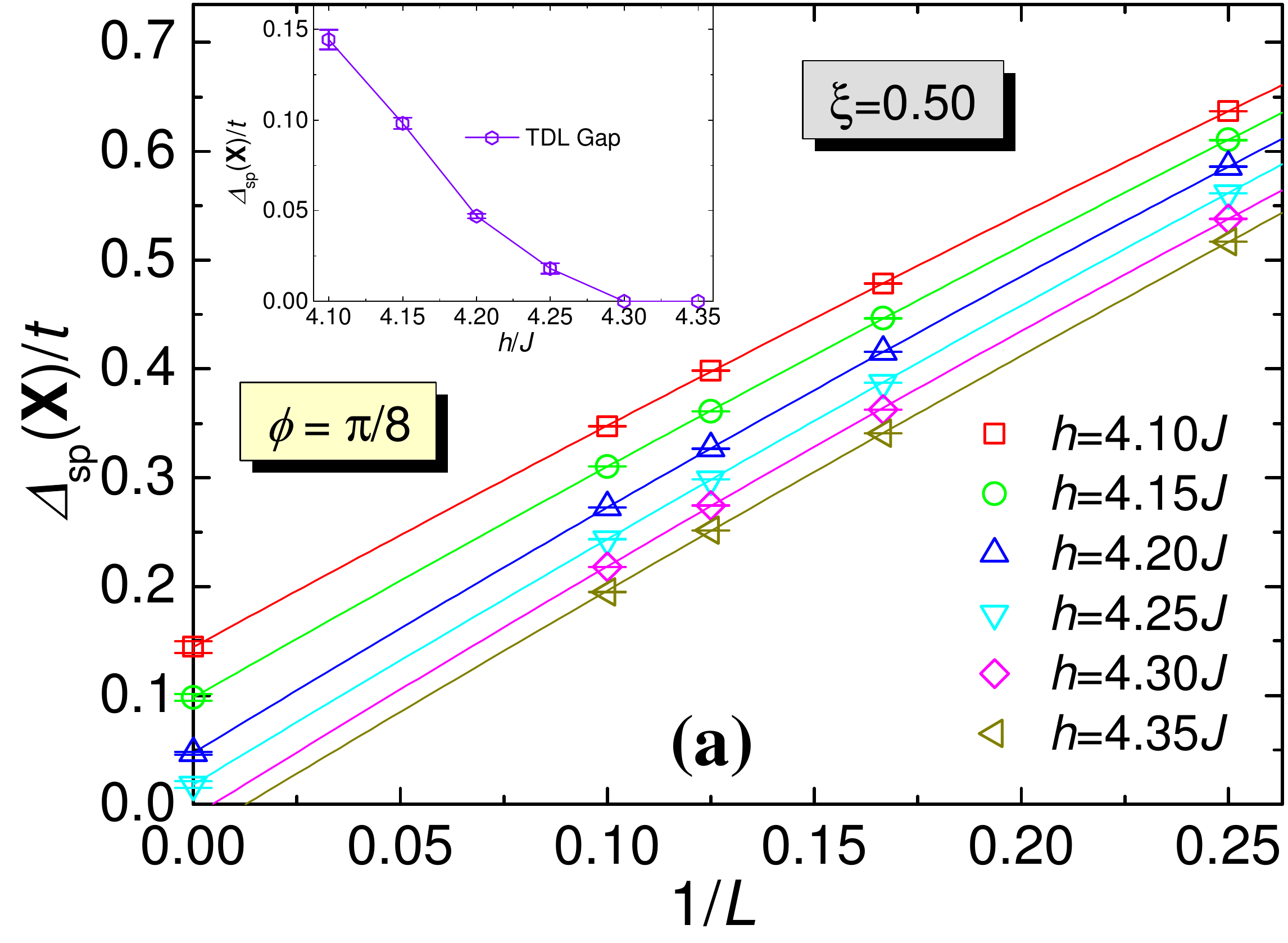}
\hspace{0.5cm}
\includegraphics[width=0.43\columnwidth]{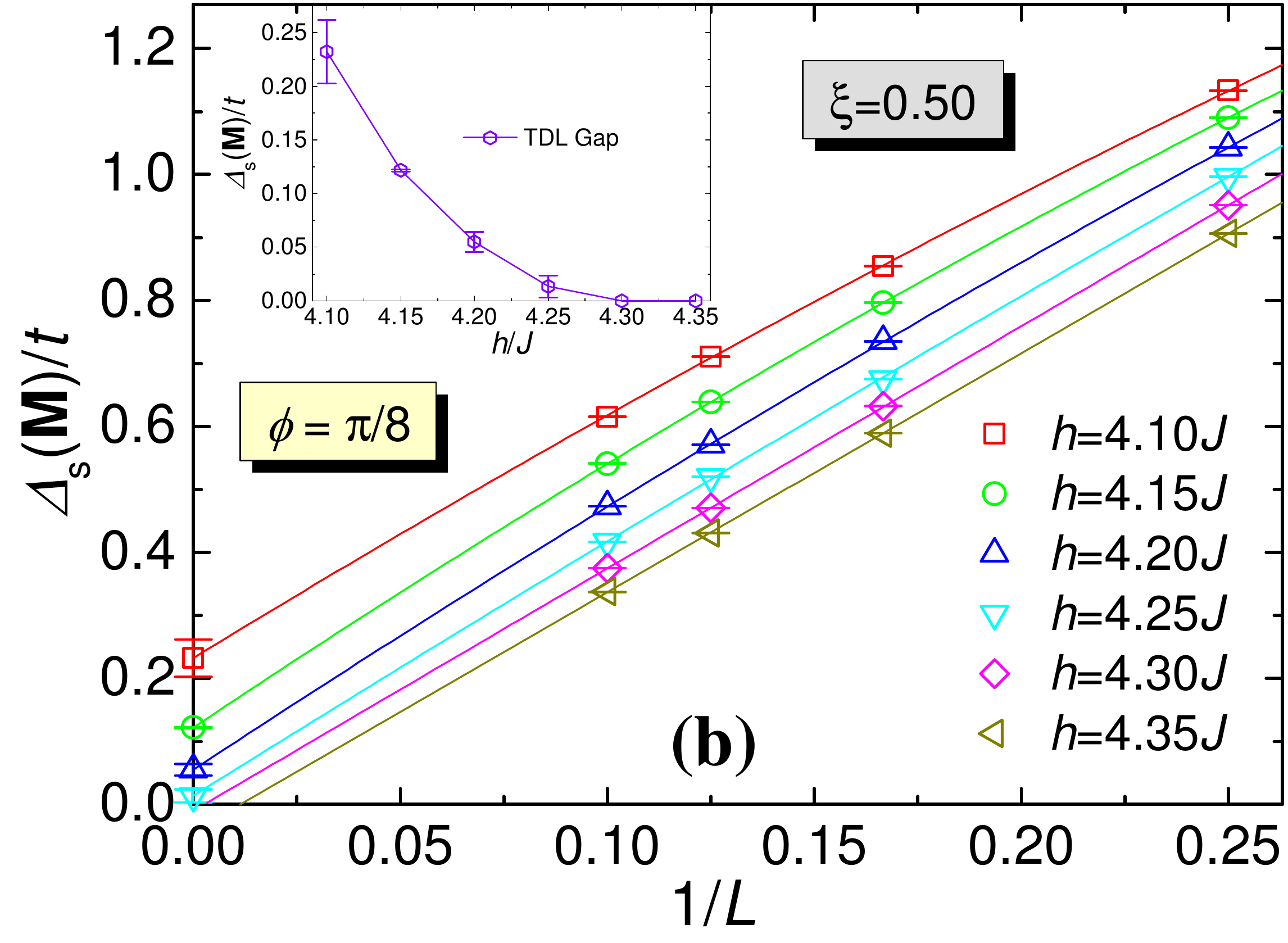}
\caption{\label{fig:Xi050HalfPiGaps} Extrapolations of (a) Single-particle Gap $\Delta_{sp}(\mathbf{X})/t$ and (b) spin gap $\Delta_s(\boldsymbol{\Gamma})/t$ over $1/L$ by second-order polynomial for $\xi=0.50$ and $\phi=\pi/8$, with $L=4,6,8,10$, case across the QCP. }
\end{figure}
The extrapolations of $\Delta_{sp}(\mathbf{X})/t$ and $\Delta_s(\mathbf{M})/t$ over $1/L$ are shown in Fig.~\ref{fig:Xi050HalfPiGaps}. We can observe that with decreasing $h/J$, both $\Delta_{sp}(\mathbf{X})/t$ and $\Delta_s(\mathbf{M})/t$ have gap opening at $h/J\in[4.25,4.30]$, suggesting the DSM-TMI phase transition at $h_c/J\in[4.25,4.30]$. This location of the QCP is consistent with the results of Binder cumulant $U_2$ and Correlation Ratio $R_{\text{Corr}}$ shown in Fig.~\ref{fig:RatioXi050HalfPi}, and also the QCP from the data collapse presented in Fig.~\ref{fig:Xi050HalfPi}. These consistency all suggest that the PM-FM phase transition for Ising spins and DSM-TMI phase transition for fermions happens simultaneously, and in the  $N=8$ Chiral Ising universality class.

\section{IV. Raw Data of dynamic properties and Excitation Gaps}
\label{sec:DynamicRawData}

In this section, we present raw data on dynamic properties, including the dynamic single-particle Green's function $G(\mathbf{X},\tau)$ and dynamic spin-spin correlation function $S^{xy}(\mathbf{M},\tau)$ in Sec. IV. A. and the comparisons of single-particle gap $\Delta_{sp}(\mathbf{X})/t$ and spin gap $\Delta_s(\mathbf{M})/t$ in Sec. IV. B.

\subsection{A. Dynamic single-particle Green's function and spin-spin correlation function}
\label{sec:SingleSpinRawData}

Since we extract the excitation gaps $\Delta_{sp}(\mathbf{X})/t$ and $\Delta_s(\mathbf{M})/t$ from $G(\mathbf{X},\tau)\propto e^{-\Delta_{sp}(\mathbf{X})\tau}$ and $S^{xy}(\mathbf{M},\tau)\propto e^{-\Delta_{s}(\mathbf{M})\tau}$ at large $\tau$ limit, we need to ensure that the data of $G(\mathbf{X},\tau)$ and $S^{xy}(\mathbf{M},\tau)$ have high quality. Here, we want to demonstrate that the dynamic data obtained from our simulations indeed have very good quality.
\begin{figure}[h!]
\centering
\includegraphics[width=0.43\columnwidth]{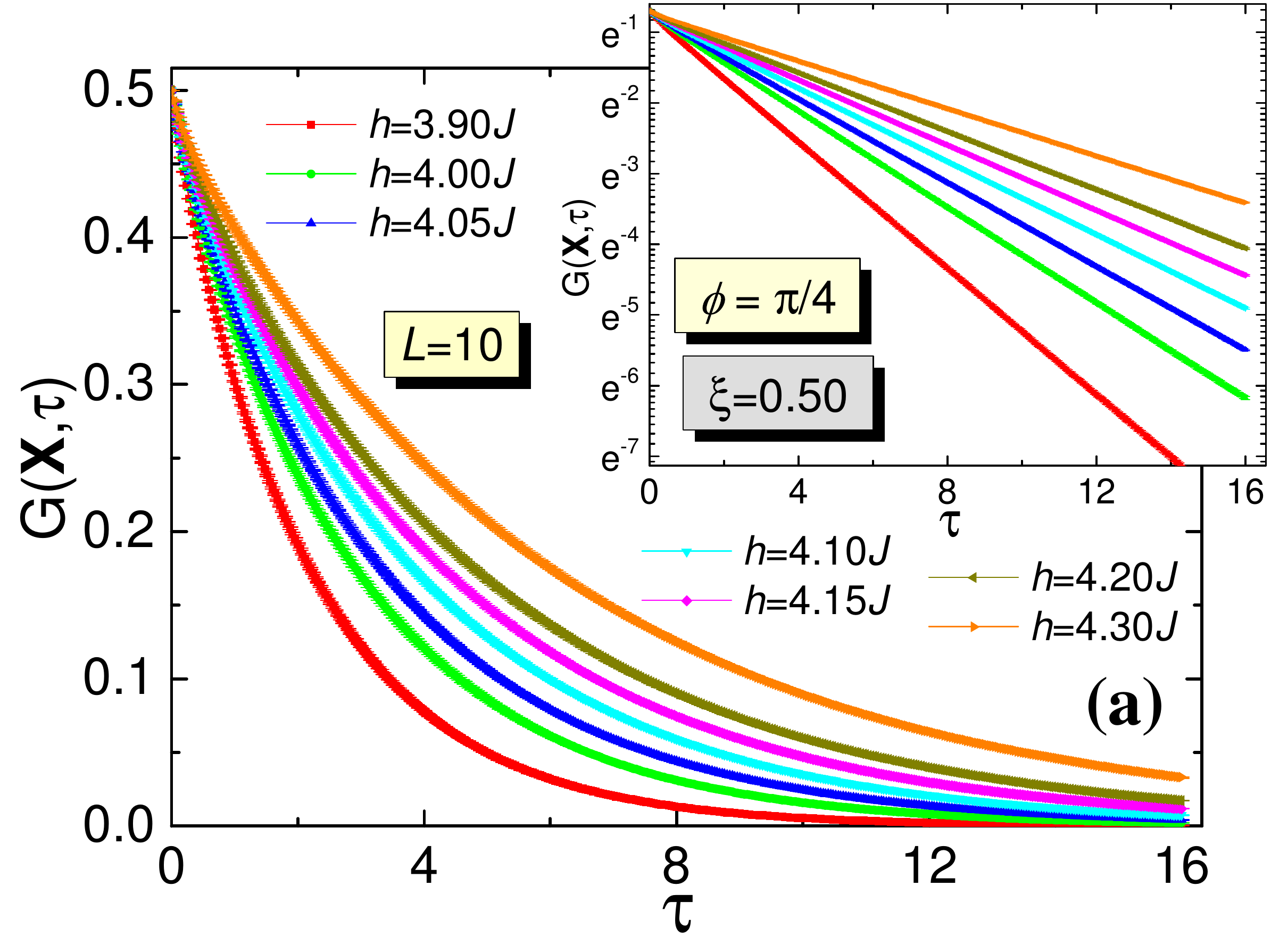}
\hspace{0.5cm}
\includegraphics[width=0.43\columnwidth]{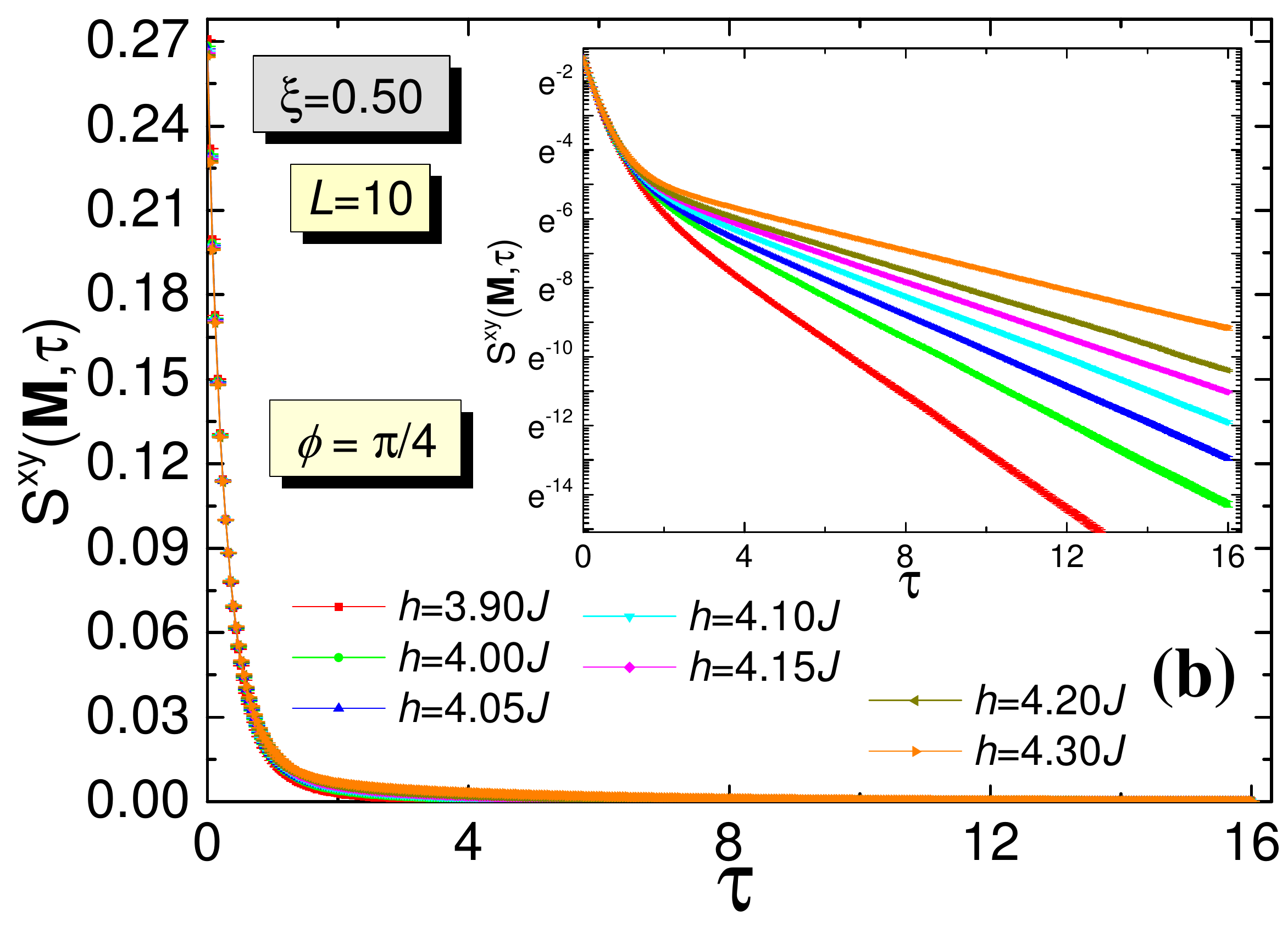}
\caption{\label{fig:Xi050PiRawData} Raw data of (a) dynamic single-particle Green's function $G(\mathbf{X},\tau)$ and (b) dynamic spin-spin correlation function $S^{xy}(\mathbf{M},\tau)$ across the DSM-TMI phase transition for $\xi=0.50$ and $\phi=\pi/4$, $L=10$. The insets are the same data plotted with semi-log coordinate, which shows the perfect linear decreasing with increasing $\tau$.}
\end{figure}

The raw data of $G(\mathbf{X},\tau)$ and $S^{xy}(\mathbf{M},\tau)$ for $\xi=0.50,\phi=\pi/4$ and $L=10$ are shown in Fig.~\ref{fig:Xi050PiRawData} with both linear and semi-log coordinates. The perfect straight lines of $\ln[G(\mathbf{X},\tau)]$ and $\ln[S^{xy}(\mathbf{M},\tau)]$ over $\tau$ allow us to extract the gaps $\Delta_{sp}(\mathbf{X})/t$ and $\Delta_s(\mathbf{M})/t$ with high precision. This good quality of the dynamic data also holds for different system size $L$, different $\xi$ parameter and also different $\phi$ such as $\phi=\pi/8$.

Notice that here we have calculated the single-particle gap at $\mathbf{X}_1$ and $\mathbf{X}_2$ points in BZ and the spin gap at $\mathbf{M}$ point. Since in the large $h$ limit, the fermions has DSM ground state and the Dirac cones locate at $\mathbf{X}_1$ and $\mathbf{X}_2$ points. With decreasing $h$, the fermions open gaps at $\mathbf{X}_1$ and $\mathbf{X}_2$ points and enter into the TMI phase. As for the spin gap, it has smallest value at $\mathbf{M}$ point, which can be understood as follows: since the coupling term $\hat{H}_{\text{Coupling}}$ in the model Hamiltonian of Eq.~(\ref{eq:ModelHamiltonianSup}) couples the Ising spin with next-nearest-neighbor (NNN) hopping of fermions, it effectively induces a NNN density-density interaction of fermions when integrate out the Ising spins, and that NNN density-density interaction favors collinear order on square lattice, which has ordered wave vector $\mathbf{Q}=\mathbf{M}$ on checkerboard lattice. We note, that this $\mathbf{Q}=\mathbf{M}$ spin order of the fermions should not be confused with the usual antiferromagnetic long-range order on the square lattice.

\subsection{B. Excitation Gaps}
\label{sec:ExcitationGaps}

In Fig.~\ref{fig:ExcitationGapsXi050} in the main text, the extrapolations of $\Delta_{sp}(\mathbf{X})/t$ and $\Delta_s(\mathbf{M})/t$ over $1/L$ are presented, and one can observe that the values of $2\times\Delta_{sp}(\mathbf{X})/t$ seem to be very close to those of $\Delta_s(\mathbf{M})/t$ within the same model parameters and system size. It's well known that the two-particle excitation gaps are twice of single-particle gap in noninteracting fermion systems. What's more, we have indeed observed $\Delta_s(\mathbf{M})/t\approx 2\times\Delta_{sp}(\mathbf{X})/t$ at small $h$, since it's very close to noninteracting fermion system. However, close to the QCP in the phase diagram, this relation does not hold any longer.

\begin{figure}[h!]
\centering
\includegraphics[width=0.55\columnwidth]{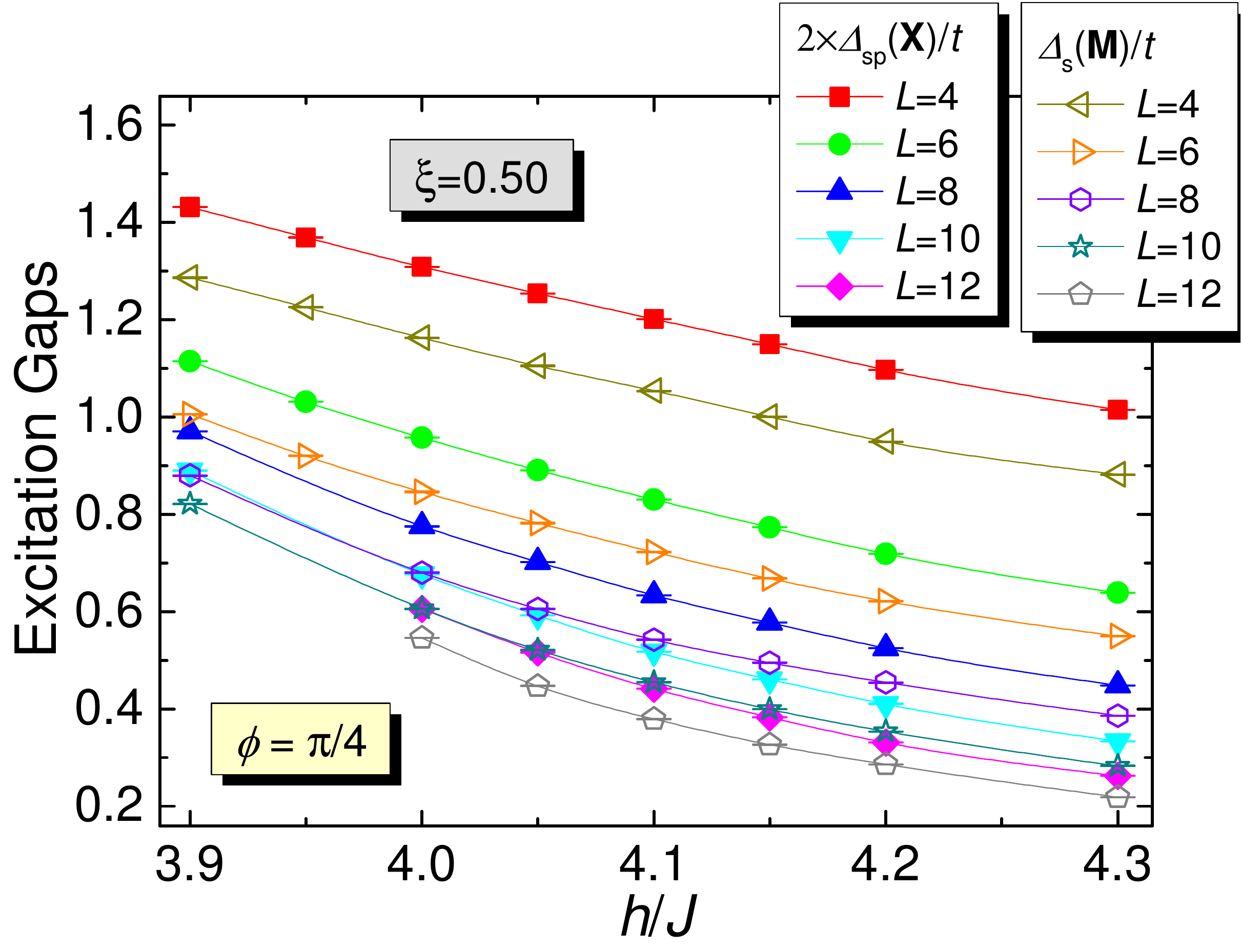}
\caption{\label{fig:Xi050PiGapsComparison}Comparison of twice of single-particle gap $2\times\Delta_{sp}(\mathbf{X})/t$ and spin gap $\Delta_s(\mathbf{M})/t$ for $\xi=0.50$ under $\pi$-flux case ($\phi=\pi/4$), for systems $L=4,6,8,10,12$, across the DSM-TMI phase transition point $h_c/J=4.11(1)$.}
\end{figure}
\begin{figure}[h!]
\centering
\includegraphics[width=0.32\columnwidth]{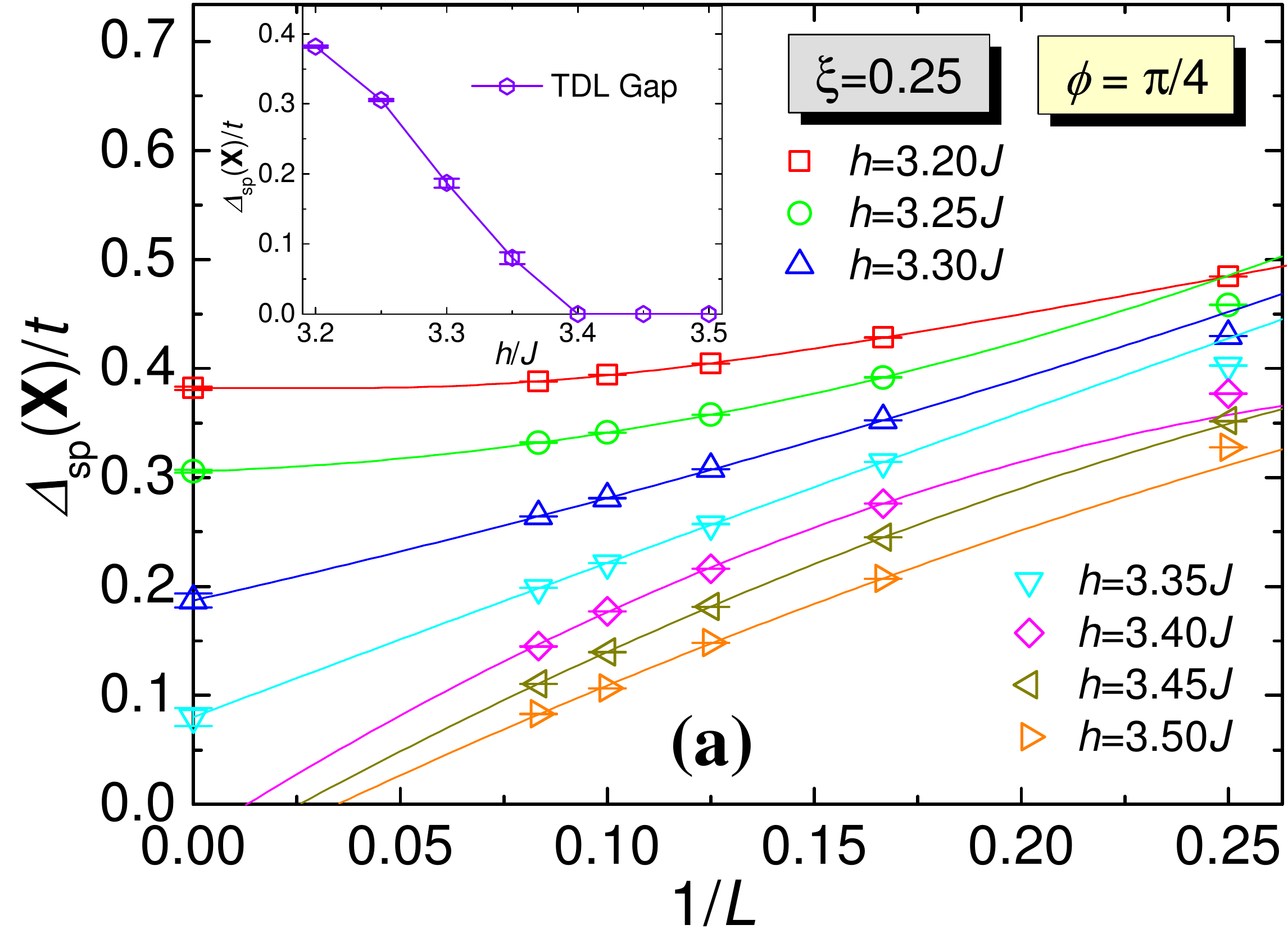}
\hspace{0.05cm}
\includegraphics[width=0.32\columnwidth]{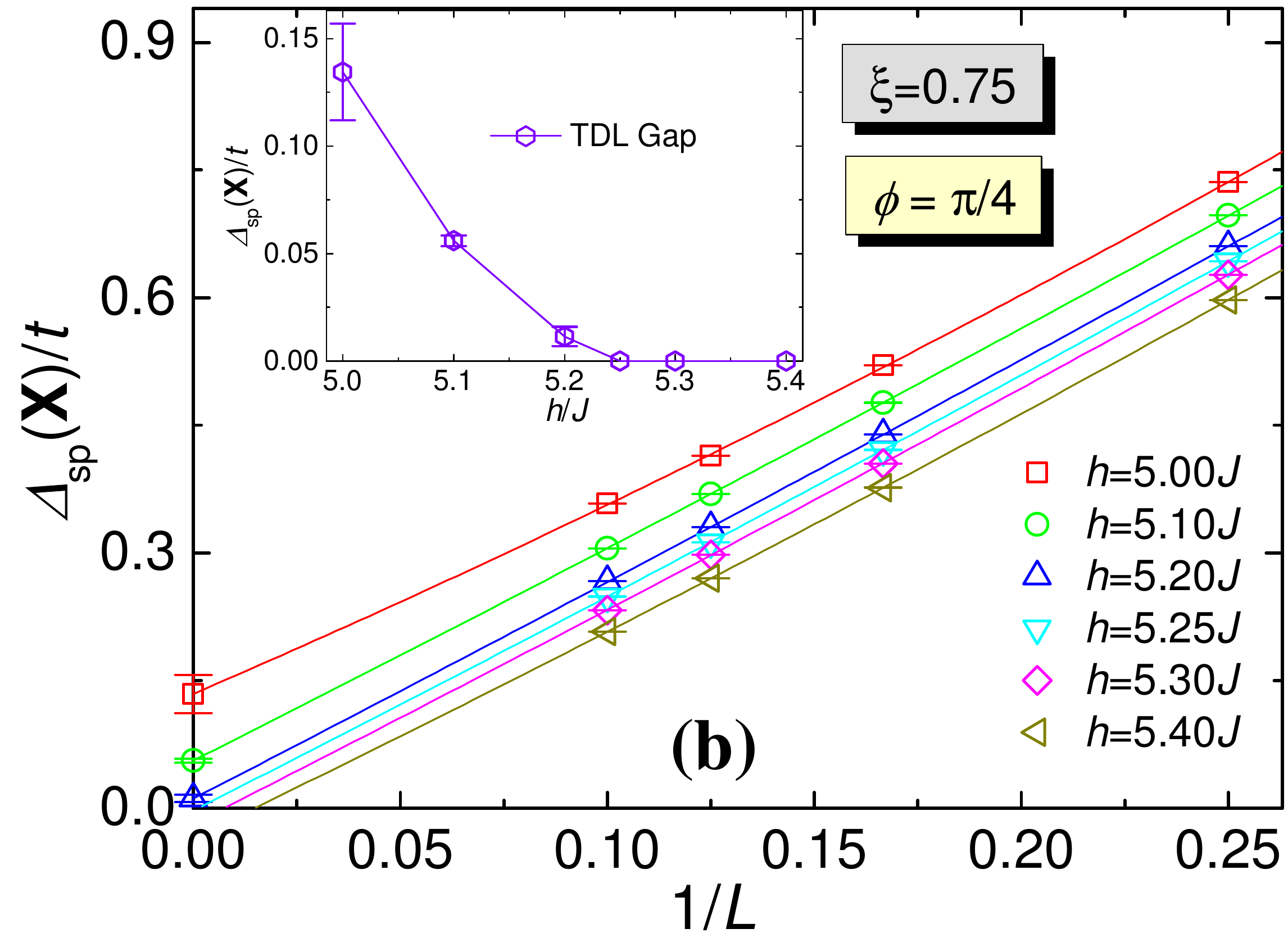}
\hspace{0.05cm}
\includegraphics[width=0.32\columnwidth]{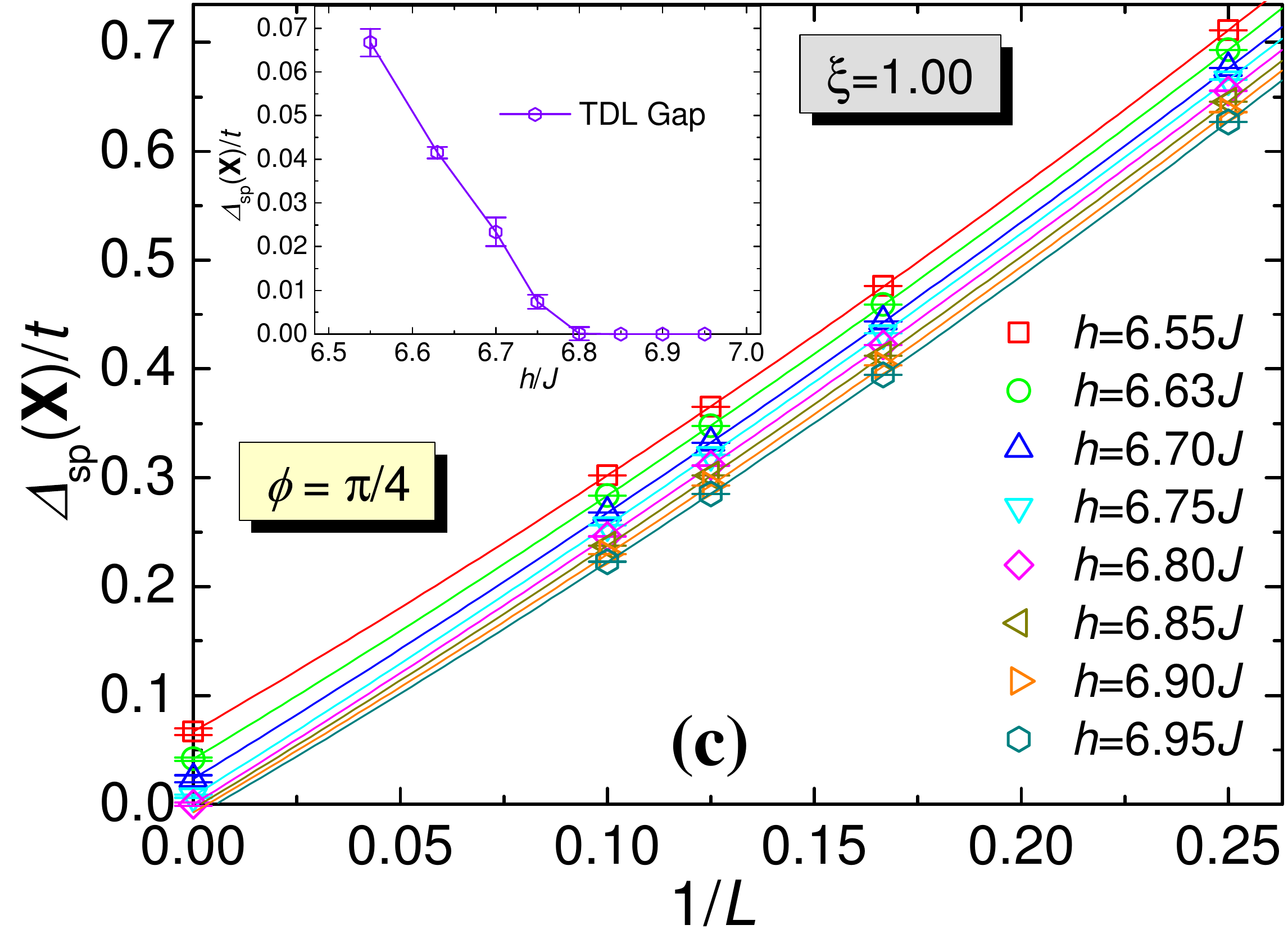}
\caption{\label{fig:Xi025075100Extrap}Extrapolations of single-particle gaps across the DSM-TMI phase transition for (a) $\xi=0.25$, (b) $\xi=0.75$ and (c) $\xi=1.00$ for $\pi$-flux case ($\phi=\pi/4$).}
\end{figure}

In Fig.~\ref{fig:Xi050PiGapsComparison}, the comparison of twice of single-particle gap $2\times\Delta_{sp}(\mathbf{X})/t$ and spin gap $\Delta_s(\mathbf{M})/t$ for $\xi=0.50$ under $\pi$-flux case ($\phi=\pi/4$), for systems $L=4,6,8,10,12$ across the DSM-TMI phase transition are shown. One can see that $2\times\Delta_{sp}(\mathbf{X})/t$ is larger than $\Delta_s(\mathbf{M})/t$. This indicates the presence of effective electron-electron interactions, mediated by the Ising spin flucutations in the model Hamiltonian. The observation that $2\times\Delta_{sp}(\mathbf{X})/t$ is larger than $\Delta_s(\mathbf{M})/t$ also holds for different $\xi$ parameter and $\phi=\pi/8$ cases close to the DSM-TMI transition.

In Fig.~\ref{fig:DataCollapseXi050} in the main text, we have shown the extrapolations of excitation gaps $\Delta_{sp}(\mathbf{X})/t$ and $\Delta_s(\mathbf{M})/t$ over $1/L$ as the proof of DSM-TMI phase transition for fermions only for $\xi=0.50$ with $\phi=\pi/4$. Here, we also present the extrapolations of $\Delta_{sp}(\mathbf{X})/t$ for $\xi=0.25,0.75,1.00$ with $\phi=\pi/4$ and show the gap opening during the DSM-TMI phase transition. The results are shown in Fig.~\ref{fig:Xi025075100Extrap}. We only present the data of $\Delta_{sp}(\mathbf{X})/t$ while the spin gap $\Delta_s(\mathbf{M})/t$ has almost the same behaviour of gap opening with decreasing $h$. For $\xi=0.25,0.75,1.00$ cases with $\phi=\pi/4$, the excitation gaps opens at $h_c/J\in[3.35,3.40]$, $h_c/J\in[5.20,5.25]$ and $h_c/J\in[6.75,6.80]$, respectively, which are well consistent with data crossing points of Binder cumulant and correlation ratio for Ising spins in Fig.~\ref{fig:RatioXi025}, Fig.~\ref{fig:RatioXi075} and Fig.~\ref{fig:RatioXi100}.

\section{V. Topological nature of TMI}
\label{sec:QSHITightBind}

In addition to the opening of excitation gaps, in this section we demonstrate the fermions in TMI indeed have QSHI ground state at $h<h_c$. At $h=0$ limit, Ising spins are classically ordered and the fermions become non-interacting, here one can analytically show the system has QSHI ground state, this part is presented in Sec. V. A. On the other hand, at finite $h<h_c$, we can also provide theoretical arguments and more importantly numerical evidence that the system is in QSHI ground state as well, demonstrated in Sec. V. B.

\begin{figure}[h!]
\centering
\includegraphics[width=0.75\columnwidth]{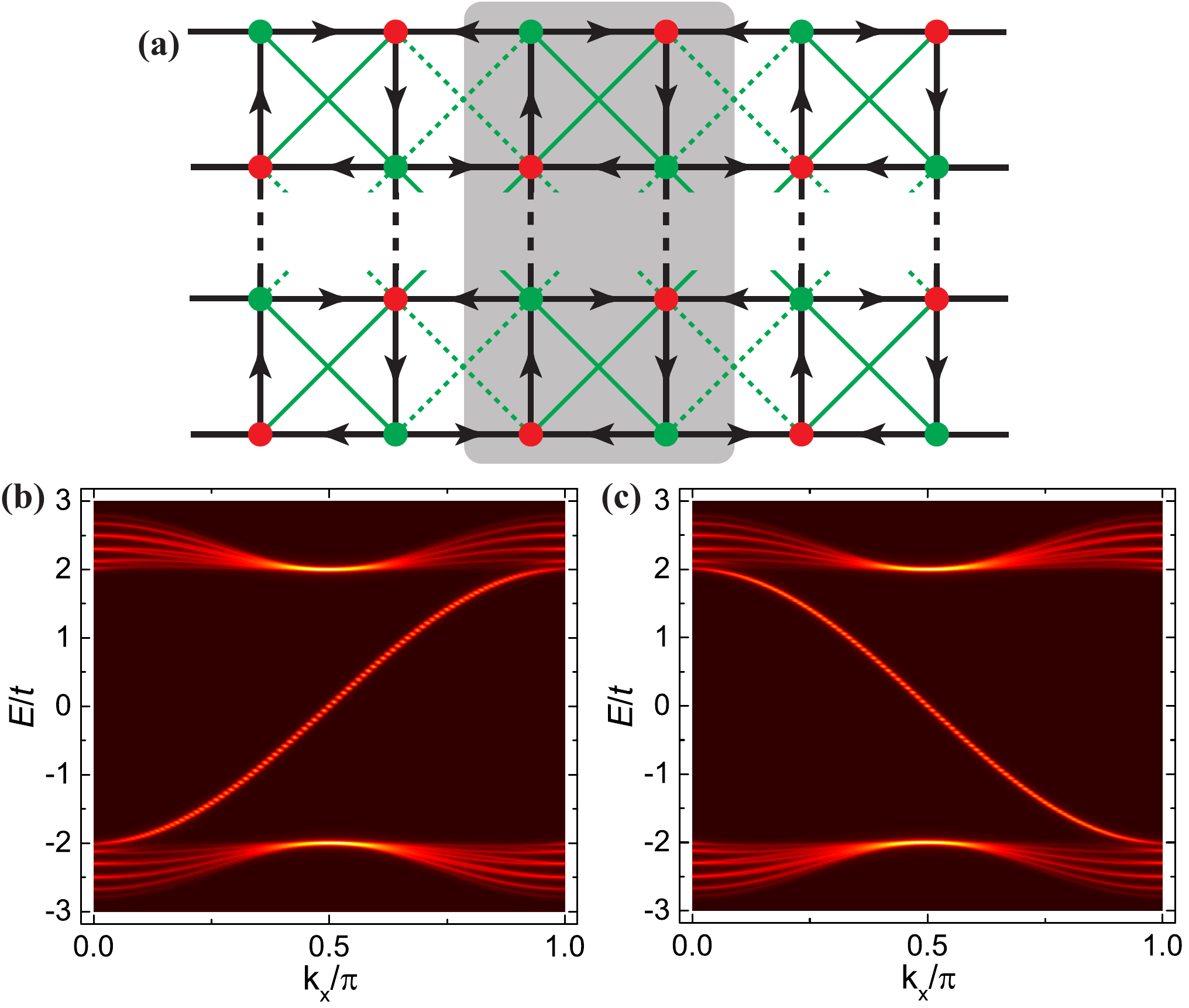}
\caption{\label{fig:h0TightBindEdgeStates}(a) Ribbon geometry for the model in Eq.~(\ref{eq:FermionModelHamt}), periodic boundary condition in $x$-direction and open boundary condition in $y$-direction. The spectral functions along one of the edges for (b) spin-up and (c) spin-down parts explicitly show the helical edge states.}
\end{figure}

\subsection{A. $h=0$ Limit}
\label{sec:h0LimitTightBind}

At $h=0$, the Ising spin in Eq.~(\ref{eq:ModelHamiltonianSup}) ordered in a classical way -- ferromagnetic order without quantum fluctuation. Thus, Hamiltonian becomes non-interacting and remaining fermion part can be written as
\begin{eqnarray}
\label{eq:FermionModelHamt}
\hat{H}_{\text{Fermion}} = -t\sum_{\langle ij \rangle\sigma} ( e^{+i\sigma\phi}c_{i\sigma}^{\dagger}c_{j\sigma} + e^{-i\sigma\phi}c_{j\sigma}^{\dagger}c_{i\sigma} )  +
 \sum_{\langle\langle ij \rangle\rangle\sigma} t_{ij}^{\prime}\xi( c_{i\sigma}^{\dagger}c_{j\sigma} + c_{j\sigma}^{\dagger}c_{i\sigma}).
\end{eqnarray}
With the choices of $t_{ij}^{\prime}=\pm t_2$, this simple tight-binding model has QSHI ground state~\cite{Sun2009,Hou2013,HanQing2016B}. This model has $U(1)_{\text{spin}}\times U(1)_{\text{charge}}\rtimes Z_2^T$ symmetry (here $Z_2^T$ stands for time-reversal symmetry), which renders the $Z$ classification of the topological index. In Fig.~\ref{fig:h0TightBindEdgeStates}, we present the spectral functions for spin-up and spin-down part along one of the edges for the model in Eq.~(\ref{eq:FermionModelHamt}) on a ribbon geometry (periodic boundary condition in $x$-direction, open boundary condition in $y$-direction) shown in Fig.~\ref{fig:h0TightBindEdgeStates}(a). We can observe that it has helical edge states with a large band gap in the bulk. We note here the topological invariant, spin Chern Number $C_s=(C_{\uparrow}-C_{\downarrow})/2=+1$, can also be calculated via the standard zero-frequency Green's function formalism~\cite{YuaoYao2016B}. Hence, combining these information, one can see that at $h=0$ the system is indeed in  a QSHI ground state for any finite $\xi$.

\subsection{B. Finite $h$ with $h<h_c$}
\label{sec:FinitehQSHI}

As shown both in the main text as well as in Sec. III. B and Sec. VI. A, the fermions in the coupling model Eq.~(\ref{eq:ModelHamiltonianSup}) has DSM ground state for $h>h_c$, namely the Ising spins in paramagnetic state. Decreasing $h$ to $h<h_c$, we have observed the opening of single-particle and two-particle excitation gaps, the system enters TMI phase with QSHI character. But since here fermions are interacting, mediated by quantum fluctuations of the Ising spins, we cannot obtain the band structure analytically as in Sec. V. A., hence will relie on the QMC results.

To numerically verify the QSHI ground state of fermions in $h<h_c$ with finite $\xi$, we calculate the topological invariant: the spin Chern number $C_s=(C_{\uparrow}-C_{\downarrow})/2$ with $C_{\sigma}$ as the Chern number for spin-$\sigma$ channel. Due to the time-reversal symmetry, $C_{\uparrow}=-C_{\downarrow}$, thus $C_s=C_{\uparrow}$. We have applied the method with zero-frequency single-particle Green's function to calculate $C_s$, which was successfully demonstrated by some of us in Ref.~\onlinecite{YuaoYao2016B}. As mentioned in the main text, since the coupling model Eq.~\ref{eq:ModelHamiltonianMain} has $Z_2$ symmetry, we need to add a pinning field $\hat{H_z}=B_z\sum_{p}s_p^z$ to Ising spins to break this $Z_2$ symmetry, incorporating the effect of spontaneously $Z_2$ symmetry breaking in thermodynamic limit of TMI phase. In practical simulations, we choose $B_z=0.001J$. As for the calculation of $C_s$, we first obtain the $\mathbf{G}_{\sigma}(\tau,\mathbf{k})$ data with both $\tau\ge0$ and $\tau<0$ from the QMC simulations as
\begin{eqnarray}
\label{eq:GreenFunction}
&& [\mathbf{G}_{\sigma}(\tau,\mathbf{k})]_{pq} = -\langle T_{\tau}[c_{\mathbf{k}p\sigma}(\tau)c_{\mathbf{k}q\sigma}^\dagger(0)] \rangle = - \frac{1}{N}\sum_{i,j=1}^N e^{-i\mathbf{k}\cdot(\mathbf{R}_i-\mathbf{R}_j)}\langle T_{\tau}[c_{ip\sigma}(\tau)c_{jq\sigma}^\dagger(0)] \rangle   \\  \nonumber
&& \tau>0 \hspace{0.4cm} \to \hspace{0.4cm}
    [\mathbf{G}_{\sigma}^{>}(\tau,\mathbf{k})]_{pq} = - \frac{1}{N}\sum_{i,j=1}^N e^{-i\mathbf{k}\cdot(\mathbf{R}_i-\mathbf{R}_j)}\langle c_{ip\sigma}(\tau)c_{jq\sigma}^\dagger \rangle    \\  \nonumber
&& \tau<0 \hspace{0.4cm} \to \hspace{0.4cm}
    [\mathbf{G}_{\sigma}^{<}(\tau,\mathbf{k})]_{pq} = + \frac{1}{N}\sum_{i,j=1}^N e^{-i\mathbf{k}\cdot(\mathbf{R}_i-\mathbf{R}_j)}\langle c_{jq\sigma}^\dagger(-\tau)c_{ip\sigma} \rangle
\end{eqnarray}
where $p,q=A,B$ standing for sublattice indexes. Then we can obtain the zero-frequency single-particle Green's function $\mathbf{G}_{\sigma}(i\omega=0,\mathbf{k})$ as
\begin{eqnarray}
\label{eq:ZeroFrequGrF}
\mathbf{G}_{\sigma}(i\omega=0,\mathbf{k}) = \int_{-\infty}^{+\infty} \mathbf{G}_{\sigma}(\tau,\mathbf{k}) d\tau = \int_{0}^{+\infty} \Big[\mathbf{G}_{\sigma}^{>}(\tau,\mathbf{k})+\mathbf{G}_{\sigma}^{<}(\tau,\mathbf{k})\Big] d\tau
\simeq \int_{0}^{+\theta} \Big[\mathbf{G}_{\sigma}^{>}(\tau,\mathbf{k})+\mathbf{G}_{\sigma}^{<}(\tau,\mathbf{k})\Big] d\tau.
\end{eqnarray}
In the last step of Eq.~(\ref{eq:ZeroFrequGrF}), a cut-off is applied to the integral, due to exponential decaying of $\mathbf{G}_{\sigma}(\tau,\mathbf{k})$ at large $\tau$. Then we can use the following formula with $\mathbf{G}_{\sigma}(i\omega=0,\mathbf{k})$ to calculate $C_{\sigma}$ for finite-size system as
\begin{eqnarray}
\label{eq:ChernProject} \mathcal{C}=&&\frac{1}{2\pi i}\iint_{\mathbf{k}\in BZ}dk_xdk_y\cdot\textrm{Tr}\Big\{P(\mathbf{k})\left[\partial_{k_x}P(\mathbf{k})\partial_{k_y}P(\mathbf{k})-\partial_{k_y}P(\mathbf{k})\partial_{k_x}P(\mathbf{k})\right]\Big\},
\end{eqnarray}
where $P(\mathbf{k})$ is a projection operator matrix constructed from eigenvectors $|\phi_m(0,\mathbf{k})\rangle$ of $\mathbf{G}_{\sigma}(i\omega=0,\mathbf{k})$:
\begin{eqnarray}
\label{eq:ProjectOperator}
P(\mathbf{k})=\sum_{\mu_m>0}|\phi_m(0,\mathbf{k})\rangle \langle\phi_m(0,\mathbf{k})|,
\end{eqnarray}
and $\mu_m$ is the corresponding eigenvalue of $\mathbf{G}_{\sigma}(0,\mathbf{k})$ with eigenvector $|\phi_m(0,\mathbf{k})\rangle$.
\begin{figure}[h!]
\centering
\includegraphics[width=0.48\columnwidth]{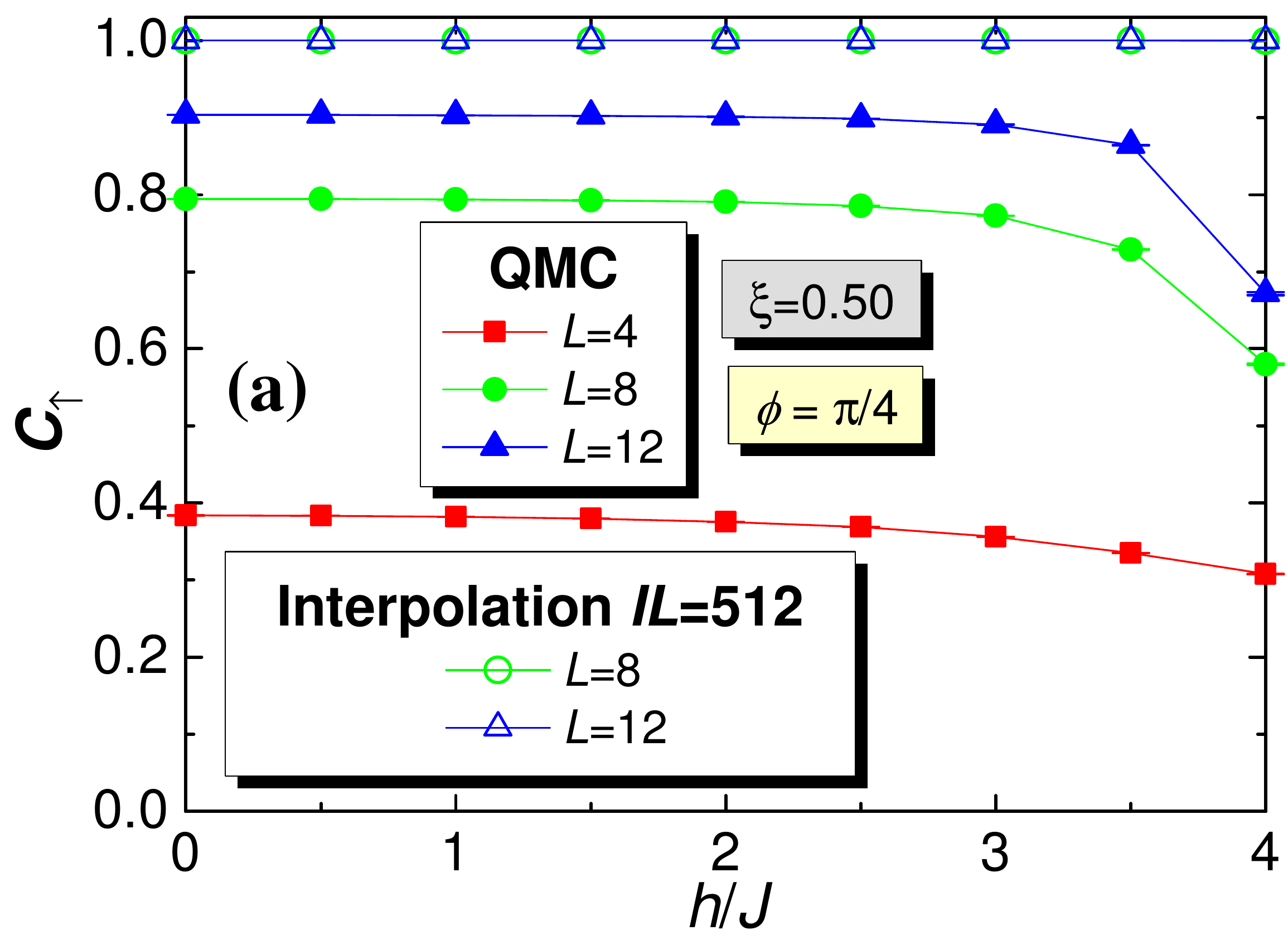}
\hspace{0.2cm}
\includegraphics[width=0.48\columnwidth]{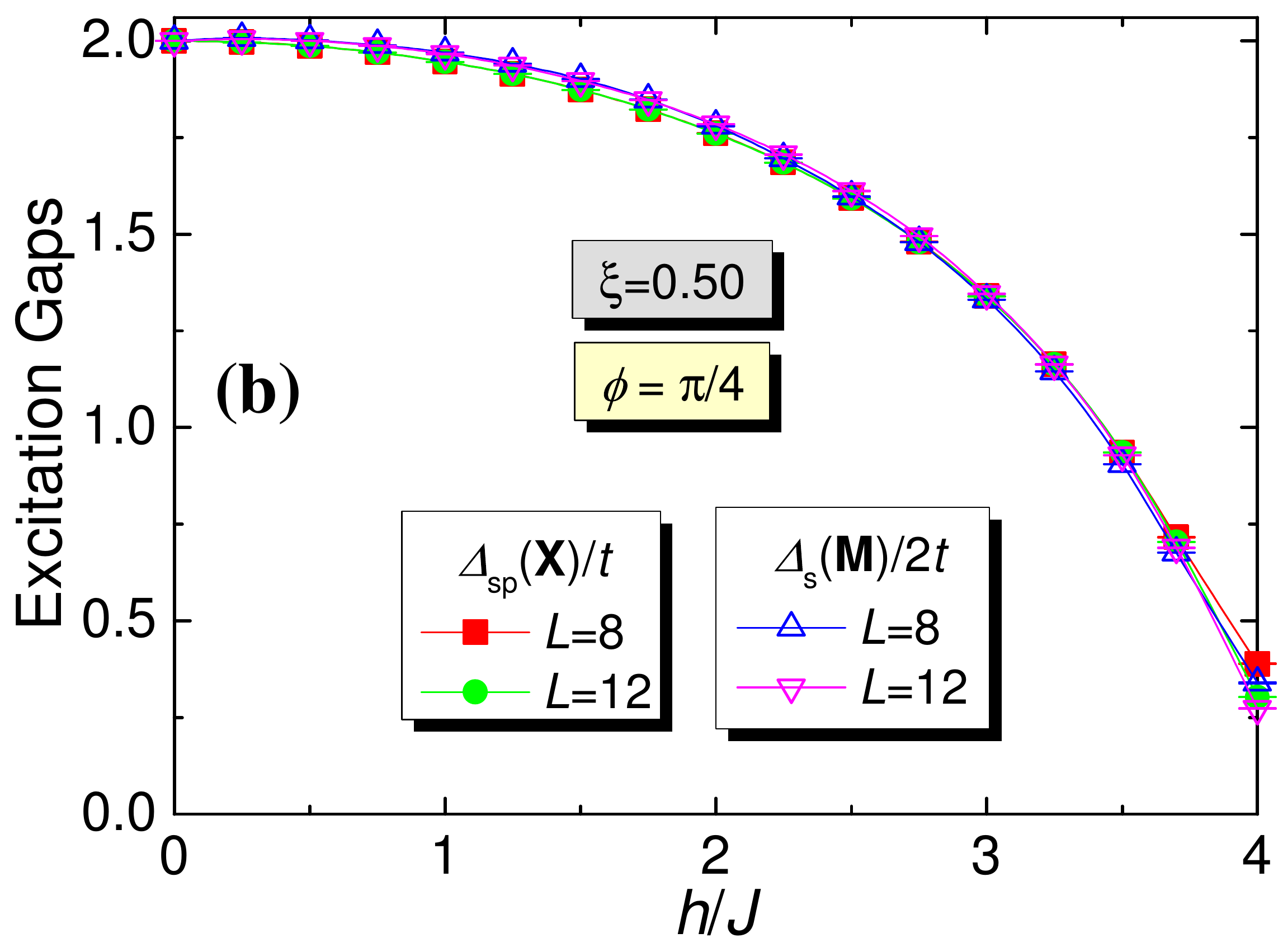}
\caption{\label{fig:ChernNumberGaps}(a) The spin Chern number $C_s$ calculated from QMC simulations of $L=4,8,12$ and also from the interpolation of $\mathbf{G}_{\sigma}(i\omega=0,\mathbf{k})$ data from $L=8,12$ systems, and (b) The single-particle gap $\Delta_{sp}(\mathbf{X})/t$ and half of spin gap as $\Delta_{s}(\mathbf{M})/2t$, in the whole region $h/J\in[0,4]$ of QSHI phase for $\xi=0.50$ and $\phi=\pi/4$. In (a), after the interpolation with system size $IL=512$, quantized integer for $C_s$ appears. }
\end{figure}
Due to the finite momentum mesh in $L\times L$ system, the spin Chern number $C_s$ calculated from Eq.~(\ref{eq:ChernProject}) suffers finite-size effect and can be away from quantized integer for small $L$. Thus, we apply the interpolation scheme of $\mathbf{G}_{\sigma}(i\omega=0,\mathbf{k})$ to achieve denser momentum resolution and approach quantized spin Chern number, as demonstrated in Ref.~\onlinecite{YuaoYao2016B}.

The results of spin Chern number $C_s$ for $L=4,8,12$ systems are shown in Fig.~\ref{fig:ChernNumberGaps}(a) for $\xi=0.50$ with $\phi=\pi/4$. Since the spin Chern number in Eq.~(\ref{eq:ChernProject}) is not well defined for gapless DSM phase, we only measure it in the QSHI phase with $h/J\in[0,4]$, i.e., $h<h_c$. As one can see, with increasing linear system size, $L=8$ to $L=12$, $C_{\uparrow}(=C_s)$ increases gradually though it's not quantized. And since $h/J=4$ is close to the QCP at $h_c=4.11(1)$ and the single-particle gap $\Delta_{sp}(\mathbf{X})/t$ is small, the cut-off $\theta$ causes the dip in $C_{\uparrow}$ though the system is still in the QSHI phase. After the interpolation with $IL=512$ from $L=8,12$ systems, $C_{\uparrow}=1$, reaches quantized integer  perfectly, meaning that the system is indeed inside TMI phase with QSHI character.

We have also measured the excitation gaps for $\xi=0.50,\phi=\pi/4$ with decreasing $h$ from $h/J=4$ to $h/J=0$, as shown in Fig.~\ref{fig:ChernNumberGaps}(b), in which we have observed that both single-particle gap $\Delta_{sp}(\mathbf{X})/t$ and spin gap $\Delta_{s}(\mathbf{M})/t$ increase monotonously to the analytical values $\Delta_{sp}(\mathbf{X})/t=2$ and $\Delta_{s}(\mathbf{M})/t=4$ at $h=0$ point. This demonstrates the fact that there is no further topological phase transition in $h<h_c$ region, the QSHI at $h=0$ smoothly cross over into TMI at finite $h$. Furthermore, we have scanned a path inside the gapped region at $h/J=3$ with $\xi\in[0,1]$ and the results for energy and structure factors for several fermion bilinears (not shown) have no signature of phase transition, suggesting that the fermions in the whole $h<h_c$ region are in TMI ground state.

\section{VI. Finite-size scaling crossover of FM-PM phase transition for Ising spins}
\label{sec:FiniteSizeScaling}

\begin{figure}[b]
\centering
\includegraphics[width=0.48\columnwidth]{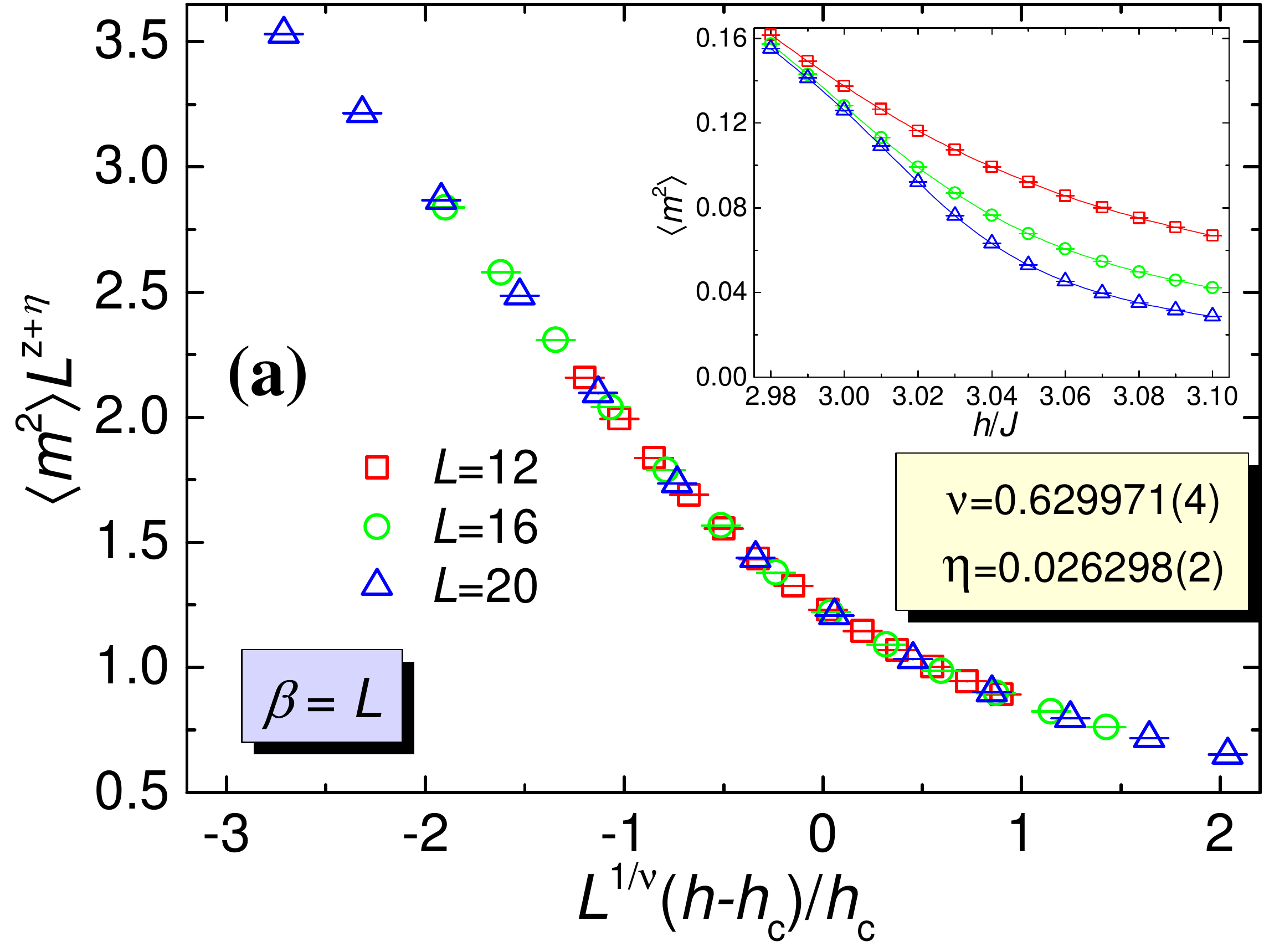}
\hspace{0.2cm}
\includegraphics[width=0.48\columnwidth]{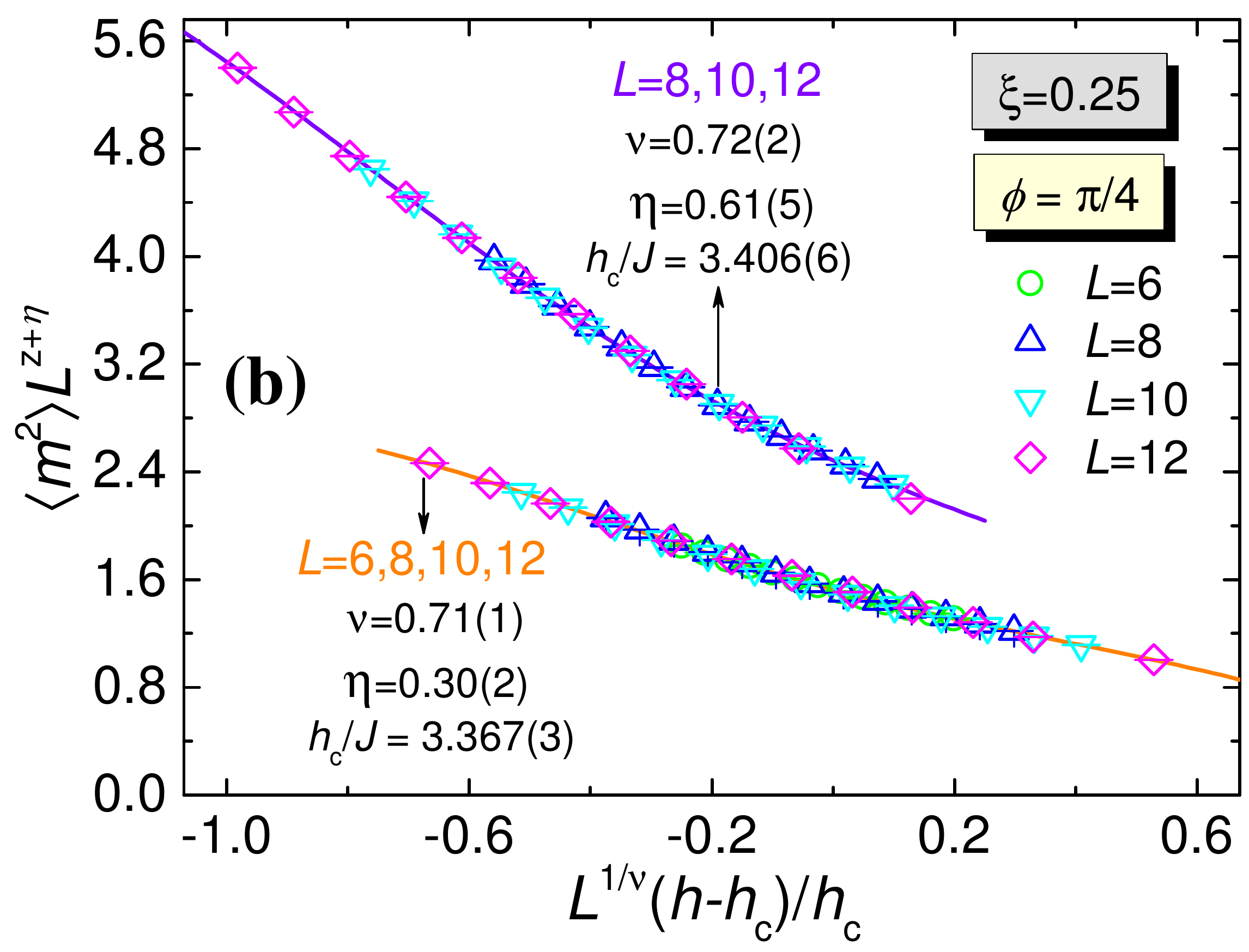}
\includegraphics[width=0.48\columnwidth]{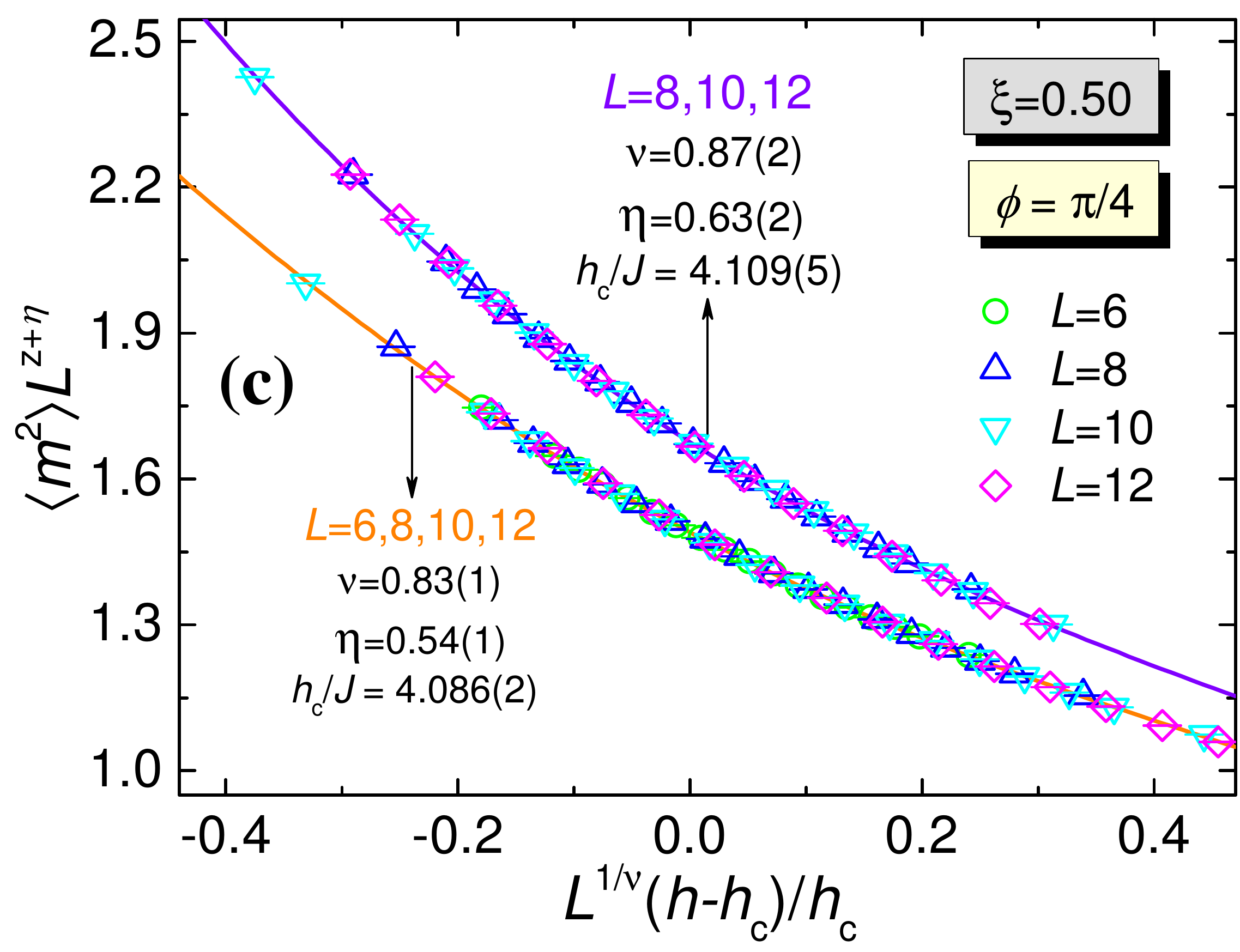}
\hspace{0.2cm}
\includegraphics[width=0.48\columnwidth]{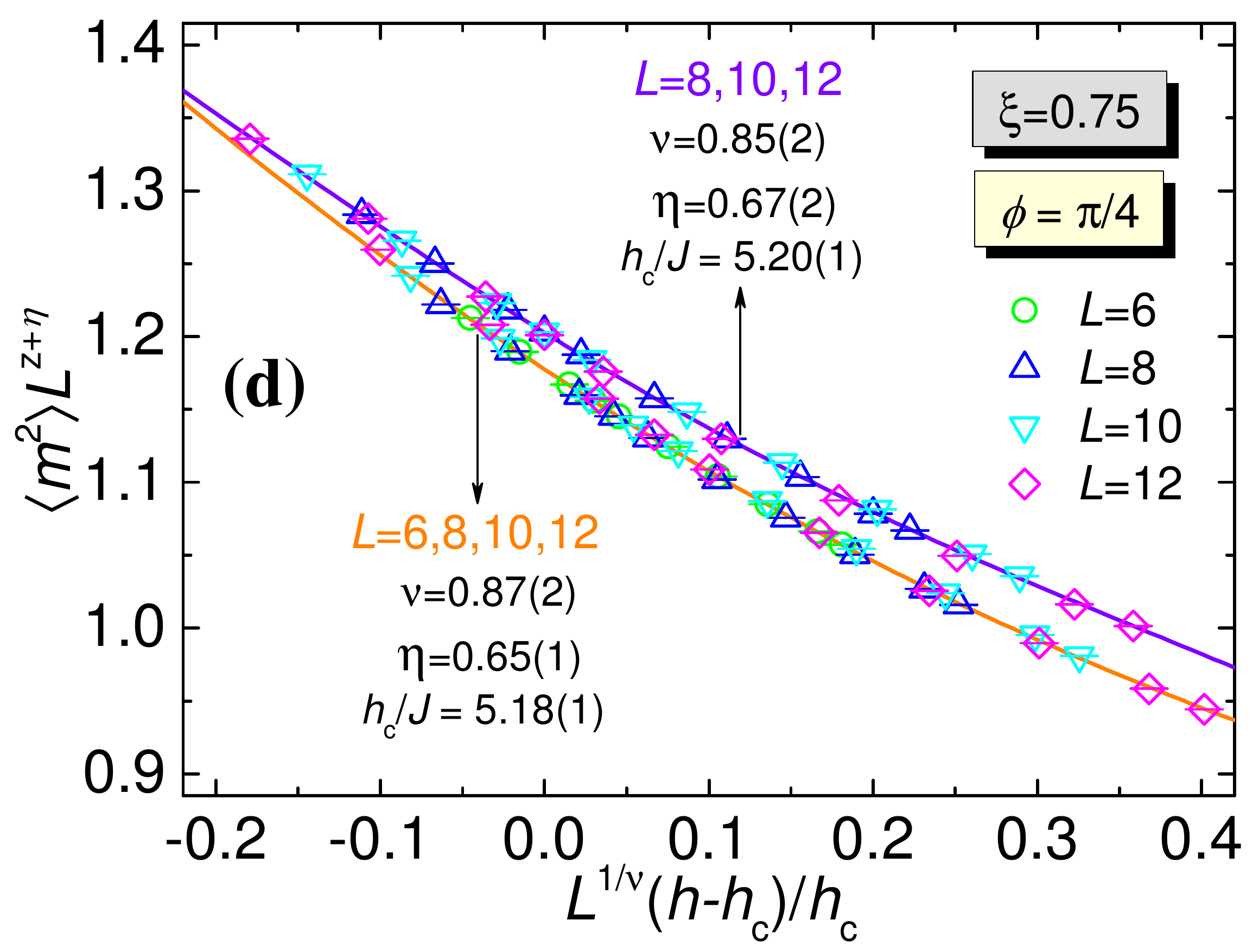}
\caption{\label{fig:ScalingCrossOver}Finite-size scaling of $\langle m^2\rangle$ data for $\xi=0.00,0.25,0.50,0.75$. (a) $\xi=0$, the collapse of $\langle m^2\rangle$ data using 3D Ising critical exponents as $\nu=0.629971(4),\eta=0.026298(2)$ with critical point $h_c/J=3.046(3)$. (b) the data collapses of $\langle m^2\rangle$ for $\xi=0.25$. (c) the data collapses of $\langle m^2\rangle$ for $\xi=0.50$. (d) the data collapses of $\langle m^2\rangle$ for $\xi=0.75$. For $\xi=0.25,0.50,0.75$, every plot contains the collapses using $\langle m^2\rangle$ data from $L=6,8,10,12$ and $L=8,10,12$, respectively.}
\end{figure}

Without coupling to fermions, 2D transverse-field Ising model has a continuous $h$-tuned quantum phase transition at 3D Ising universality class. Thus, at $\xi=0$ point of model in Eq.~(\ref{eq:ModelHamiltonianMain}), the $h$-tuned quantum phase transition belongs to the 3D Ising universality class with $\nu=0.629971(4),\eta=0.026298(2)$~\cite{Hasenbusch1998}. With coupling to fermion, however, the universality class is altered to the $N=8$ Chiral Ising. Therefore, there are two different kinds of universality classes along the phase transition line (the red line) in the ground-state phase diagram in Fig.~\ref{fig:LatticeBZPhaseDiagram} (c) in main text. A natural question here is, how the 3D Ising universality class evolves into the $N=8$ Chiral Ising universality class. In this section, we present some results on this problem.

We think that there exists a {\it{finite-size scaling crossover}} behavior of the universality class for the QCPs of Ising spins. In the thermodynamic limit, the QCP should belong to 3D Ising universality at $\xi=0$ point and $N=8$ Chiral Ising universality class for infinitesimally small $\xi$. However, in the finite-size systems simulated, we are expected to see the following features. First, at small $\xi$, the critical exponents obtained from finite-size scaling applied to data of small system sizes are closer to 3D Ising universality class, while those obtained from data of larger system sizes are closer to $N=8$ Chiral Ising universality class. Second, the finite-size scalings of $\langle m^2\rangle$ data near QCPs from the same system size for increasing $\xi$ parameters should arrive at critical exponents closer to those of $N=8$ Chiral Ising universality class. These features imply that there exist a length scale $L_c$ for each $\xi$, below which the critical exponents obtained from scaling should be close to 3D Ising universality class and otherwise close to $N=8$ Chiral Ising universality class. Clearly, the $L_c$ should be larger for smaller $\xi$ and becomes smaller for larger $\xi$.

The finite-size scalings of $\langle m^2\rangle$ data from $\xi=0.00,0.25,0.50,0.75$ are shown in Fig.~\ref{fig:ScalingCrossOver}. First of all, we have applied the critical exponents $\nu=0.629971(4),\eta=0.026298(2)$ to collapse the $\langle m^2\rangle$ data for $\xi=0$ and we have obtained $h_c/J=3.046(3)$, which is well-consistent with previous results~\cite{XuXiaoYan2017a,XuXiaoYan2017b} and the data collapse also have high quality. Then for $\xi=0.25,0.50,0.75$, we have performed the data collapses with free $\nu,\eta,h_c$ and data from $L=6,8,10,12$ and $L=8,10,12$, respectively. For $\xi=0.25$, we can observe a dramatic change of $\eta$ exponent from $\eta=0.30(2)$ to $\eta=0.61(5)$, by simply adding the $\langle m^2\rangle$ data of $L=12$ system. This fact explicitly shows that for smaller system sizes like $L=6,8$, the critical behavior is closer to 3D Ising universality, while it's more likely to be $N=8$ Chiral Ising universality class for $L=12$ system. We can conclude $L_c\approx10$ for $\xi=0.25$, signifying the {\it{finite-size scaling crossover}} behavior. The critical exponents from data collapses of $L=6,8,10,12$ and $L=8,10,12$ are both converging to the numbers in $N=8$ Chiral Ising universality class, for increasing $\xi=0.25,0.50,0.75$ parameters. This indicates the decreasing $L_c$ length scale in the finite-size scaling for increasing $\xi$. Third, we can also observe that the $\eta$ exponent suffers much stronger finite-size effect than that of $\nu$ exponent, especially for small $\xi$. This is simply due to the fact that the two universality classes have similar $\nu$ exponents while $\eta$ exponent differs a lot. As a result, the crossover between these two universality classes gives much larger deviation for $\eta$ than $\nu$. All of these numerical results support the {\it{finite-size scaling crossover}} behavior in the ground-state phase diagram.

\end{document}